\pdfoutput=1
\RequirePackage{ifpdf}
\ifpdf 
\documentclass[pdftex]{sigma}
\else
\documentclass{sigma}
\fi

\numberwithin{equation}{section}

\begin{document}

\allowdisplaybreaks

\renewcommand{\thefootnote}{$\star$}

\renewcommand{\PaperNumber}{090}

\FirstPageHeading

\ShortArticleName{The Klein--Gordon Equation and Dif\/ferential Substitutions}

\ArticleName{The Klein--Gordon Equation and Dif\/ferential\\
 Substitutions of the Form $\boldsymbol{v=\varphi(u,u_x,u_y)}$\footnote{This
paper is a contribution to the Special Issue ``Symmetries of Dif\/ferential Equations: Frames, Invariants and Applications''. The full collection is available at \href{http://www.emis.de/journals/SIGMA/SDE2012.html}{http://www.emis.de/journals/SIGMA/SDE2012.html}}}

\Author{Mariya N. KUZNETSOVA~$^\dag$, Asl{\i} PEKCAN~$^\ddag$ and Anatoliy V. ZHIBER~$^\S$}

\AuthorNameForHeading{M.N.~Kuznetsova, A.~Pekcan and A.V.~Zhiber}

\Address{$^\dag$~Ufa State Aviation Technical University, 12 K. Marx Str., Ufa, Russia}
\EmailD{\href{mailto:kuznetsova@matem.anrb.ru}{kuznetsova@matem.anrb.ru}}

\Address{$^\ddag$~Department of Mathematics, Istanbul University, Istanbul, Turkey}
\EmailD{\href{mailto:pekcan@istanbul.edu.tr}{pekcan@istanbul.edu.tr}}

\Address{$^\S$~Ufa Institute of Mathematics, Russian Academy of Science,\\
\hphantom{$^\S$}~112 Chernyshevskii Str., Ufa, Russia}
\EmailD{\href{mailto:zhiber@mail.ru}{zhiber@mail.ru}}

\ArticleDates{Received April 25, 2012, in f\/inal form November 14, 2012; Published online November 26, 2012}

\Abstract{We present the complete classif\/ication of equations of the form $u_{xy} = f(u, u_x, u_y)$
and the Klein--Gordon equations $v_{xy} = F(v)$ connected with one another
by dif\/ferential substitutions $v = \varphi(u, u_x, u_y)$ such that
$\varphi_{u_x}\varphi_{u_y}\neq 0$ over the ring of complex-valued variables.}


\Classification{35L70}

\vspace{-3mm}

\renewcommand{\thefootnote}{\arabic{footnote}}
\setcounter{footnote}{0}

\section{Introduction}

In this paper, we study the classif\/ication problem of equations of the form
\begin{gather}
u_{xy} = f(u,u_x,u_y) \label{u_xy}
\end{gather}
over the ring of complex-valued variables. Such equations have applications in many f\/ields of mathematics and physics.
Liouville~\cite{Liouville}, B\"acklund~\cite{Backl1}, Darboux~\cite{Darb} and other authors~\cite{Bianchi, Tzitz} studying the surfaces of constant negative curvature discovered the f\/irst examples of integrable nonlinear hyperbolic equations. In the 1970s, one of the fundamental methods of mathematical physics, the inverse scattering method, was introduced. After that, since hyperbolic equations have many applications in physics (continuum mechanics, quantum f\/ield theory, theory of ferromagnetic materials etc.), many important studies were published.

Existence of higher symmetries is a hallmark of integrability of an equation. Drinfel'd, Sokolov and Svinolupov~\cite{DrSvSok, Sv} showed that symmetries can be ef\/fectively used for classif\/ication of evolution equations.
Zhiber and Shabat~\cite{ZhiberShabat} obtained the
complete list of the Klein--Gordon equations
\begin{gather}
v_{xy} = F(v) \label{v_xy}
\end{gather}
with higher symmetries. However, the
symmetry method for the classif\/ication of equations of form \eqref{u_xy} faces particular dif\/f\/iculties. Therefore, here we use dif\/ferential substitutions to solve the classif\/ication problem.

Before going further, let us give some def\/initions. Let $u$ be a solution of equation~\eqref{u_xy}. All the mixed derivatives of $u$
\begin{gather} \label{kuzn_variations}
u_x, \qquad u_{y}, \qquad u_{xx}, \qquad u_{yy}, \qquad\dots
\end{gather}
will be expressed through equation~\eqref{u_xy} with dif\/ferential consequences of this equation. Here~$u$ and variables \eqref{kuzn_variations} will be regarded as independent.

We begin with an important notion of (inf\/initesimal) symmetry of equation \eqref{u_xy}. Denote the operators of total derivatives with respect to~$x$ and~$y$ by $D$ and $\bar{D}$, respectively.

\begin{definition}
The symmetry of equation~\eqref{u_xy} of order $(n,m)$ is the function $g = g(u, u_1, \dots, u_n$, $\bar{u}_1, \dots, \bar{u}_m)$, $g_{u_n} \neq 0$, $g_{\bar{u}_m} \neq 0$, satisfying the equation
\[
(D \bar{D} - f_{u_1} D - f_{\bar{u}_1} \bar{D} - f_u)g = 0.
\]
Here $u_i = \frac{\partial^i u}{ \partial x^i}$ and $\bar{u}_i = \frac{\partial^i u}{ \partial y^i}$, $i \in \mathbb{N}$. If $n \leq 1$ and $m \leq 1$ then the function $g$ is called a classical symmetry, otherwise we have a higher symmetry.
\end{definition}
Assume that $g$ is a symmetry of equation~\eqref{u_xy}. It is easy to check that the derivatives $g_{u_n}$ and $g_{\bar{u}_m}$ satisfy the so-called characteristic equations $\bar{D} (g_{u_n}) = 0$ and $D (g_{\bar{u}_m}) = 0$,
respectively. It actually can be shown that $g_{u_n}$ depends only on the variables $u, u_1, \dots, u_n$, while $g_{\bar{u}_m}$ is a~function of the variables $u, \bar{u}_1, \dots, \bar{u}_m$.

\begin{definition} The function $\omega(u, u_1, u_2, \dots, u_n)$, $\omega_{u_n} \neq 0$, is called an $x$-integral of order $n$ of equation~\eqref{u_xy} if $\bar{D}(\omega) =0$. Similarly, the $y$-integral of order $m$ is the function $\bar{\omega}(u, \bar{u}_1, \bar{u}_2, \dots$, $\bar{u}_m)$, $\bar{\omega}_{\bar{u}_m} \ne 0$ which satisf\/ies $D(\bar{\omega}) =0$.
\end{definition}

Another important notion is the sequence of the Laplace invariants of equation~\eqref{u_xy}.
\begin{definition}
The main generalized Laplace invariants of equation~\eqref{u_xy} are the functions $H_0$ and $H_1$ given by the formulae
\[
H_1 = - D \left( \frac{\partial f}{\partial u_1}\right) + \frac{\partial f}{\partial u_1} \frac{\partial f}{\partial \bar{u}_1} + \frac{\partial f}{\partial u}, \qquad
H_0 = - \bar{D} \left( \frac{\partial f}{\partial \bar{u}_1}\right) + \frac{\partial f}{\partial u_1} \frac{\partial f}{\partial \bar{u}_1} + \frac{\partial f}{\partial u}.
\]
\end{definition}

 Other Laplace invariants can be found recurring in the relation
\[
D \bar{D} (\ln H_i) = -H_{i+1} - H_{i-1} + 2H_i, \qquad i \in \mathbb{Z}.
\]
Sokolov and Zhiber~\cite{ZhSok} showed that the functions $H_1$ and $H_0$ are invariants of equation~\eqref{u_xy} under the point transformations $u \rightarrow \zeta(x,y,u)$. Generalized Laplace invariants play a signif\/icant role in the investigation of integrability of equations. Namely, Anderson and Kamran~\cite{Anderson2}, Zhiber, Sokolov and Startsev~\cite{ZhSokSt} proved that an equation has nontrivial $x$- and $y$-integrals if and only if the Laplace sequence of invariants terminates on both sides ($H_r = H_s \equiv 0$ for some values $r$ and $s$), which is indeed a def\/inition of the (Darboux) integrability of an equation. Equations satisfying the last condition are called Liouville type
equations. Using this def\/inition for linear equations $V_{xy} + a(x,y)V_x + b(x,y)V_y + c(x,y)V = 0$, one can obtain equations with the f\/inite Laplace sequence studied in detail by Goursat~\cite{Goursat}.

It should be noted that symmetries of Liouville type equations have two arbitrary functions, while the equations integrable by the inverse scattering method (for instance, the sine-Gordon equation) have a countable set of symmetries.

The main notion of the paper is the notion of dif\/ferential substitutions.

\begin{definition}
The relation
\begin{gather} \label{kuzn_subs}
v = \varphi \left( u, \frac{\partial u}{\partial x}, \dots,\frac{\partial^n u}{\partial x^n}, \frac{\partial u}{\partial y}, \dots, \frac{\partial^m u}{\partial y^m} \right)
\end{gather}
is called a dif\/ferential substitution from equation~\eqref{u_xy} to the equation
\begin{gather} \label{def_v_xy}
v_{xy} = g(v, v_x, v_y)
\end{gather}
if function \eqref{kuzn_subs} satisf\/ies equation~\eqref{def_v_xy} for every
solution $u(x,y)$ of equation~\eqref{u_xy}.
\end{definition}

Before proceeding, let us brief\/ly mention some works related to dif\/ferential substitutions.
Sokolov~\cite{Sok} showed that substitutions can be used in the study of integrability of nonlinear dif\/ferential equations.
There exist various dif\/ferent def\/initions of exact integrable hyperbolic equations. Sokolov and Zhiber~\cite{ZhSok} presented one of the most comprehensive reviews of such equations. As mentioned before, existence of higher symmetries is a hallmark of integrability of
an equation. Meshkov and Sokolov~\cite{MeshSok} presented the complete list of one-f\/ield hyperbolic equations with generalized integrable $x$- and $y$-symmetries of the third order. One can f\/ind many examples of nonlinear equations and dif\/ferential substitutions in~\cite{MeshSok, ZhSok}. Star\-tsev~\cite{St2, St1} described properties of generalized Laplace invariants of nonlinear equations with dif\/ferential substitutions. B\"acklund transformations and, in particular cases, dif\/ferential substitutions were studied by Khabirov~\cite{Kh2}. Kuznetsova~\cite{Kuzn1} described coupled equations for which linearizations are related by Laplace transformations of the f\/irst and the second orders. A~B\"acklund transformation was constructed
for such pairs.

Although we know a considerable amount of nonlinear equations which are connected with one another by dif\/ferential substitutions, the problem of classifying dif\/ferential substitutions and B\"acklund transformations was solved only for evolution equations.

Recently, Zhiber and Kuznetsova~\cite{Kuzn2} have applied dif\/ferential substitutions to classify equations. Namely, all equations of form \eqref{u_xy} are transformed into equations of form \eqref{v_xy} by dif\/ferential substitutions of the special form
$
v = \varphi(u, u_x)
$
were described. All these equations are contained in the following list:
\begin{alignat*}{4}
&u_{xy} = u F' \bigl( F^{-1} (u_x) \bigr), \qquad &&v_{xy} = F(v), \qquad &&v = F^{-1}(u_x); &\\[-0.5ex]
&u_{xy} = \sin u \sqrt{1 - u^2_x}, \qquad &&v_{xy} = \sin v, \qquad &&v = u + \arcsin u_x;& \\[-0.5ex]
&u_{xy} = \exp u \sqrt{1 + u^2_x}, \qquad &&v_{xy} = \exp v, \qquad &&v = u + \ln \left( u_x + \sqrt{1 + u^2_x} \right);&\\[-1ex]
&u_{xy} = \frac{\sqrt{2 u_y}}{s'(u_x)}, \qquad &&v_{xy} = F(v), \qquad &&v = s(u_x),&
\intertext{where the functions $s$ and $f$ satisfy $s'(u_x) F( s(u_x) ) = 1$;}
& u_{xy} = \frac{c - u_y \varphi_u(u, u_x)}{\varphi_{u_x}(u, u_x)}, \qquad && v_{xy} = 0, \qquad && v = \varphi(u, u_x); &
\\
& u_{xy} = u_x \bigl( \psi(u, u_y) - u_y \alpha'(u) \bigr), \qquad && v_{xy} = \exp v, \qquad && v = \alpha(u) + \ln u_x, &
\intertext{where $\psi_u + \psi \psi_{u_y} - \alpha' u_y \psi_{u_y} = \exp \alpha$;}
& u_{xy} = u_x \bigl( \psi(u, u_y) - u_y \alpha'(u) \bigr), \qquad && v_{xy} = 0, \qquad && v = \alpha(u) + \ln u_x,&
\intertext{where $\psi_u + \psi \psi_{u_y} - \alpha' u_y \psi_{u_y} = 0$;}
& u_{xy} = u, \qquad && v_{xy} = v, \qquad && v = c_1 u + c_2 u_x; & \\
& u_{xy} = \delta(u_y), \qquad && v_{xy} = 1, \qquad && v = c_1 u + c_2 u_x, \qquad \delta(c_1 + c_2 \delta') = 1,&
\end{alignat*}

\newpage

\noindent
up to the point transformations $u \rightarrow \theta(u)$, $v \rightarrow \kappa(v)$, $x \rightarrow \xi x$, and $y \rightarrow \eta y$, where~$\xi$ and~$\eta$ are arbitrary constants. Here $c$ is an arbitrary constant, $c_1$ and $c_2$ are constants satisfying $(c_1,c_2)\ne(0,0)$, and the function $\psi$ satisf\/ies $(\psi_u, \psi_{u_y}) \neq (0, 0)$.

Furthermore, all equations of form \eqref{v_xy} that can be transformed into equations of form \eqref{u_xy} by dif\/fe\-ren\-tial substitutions of the form $u = \psi(v, v_y)$ are given in the following list:
\begin{alignat*}{6}
&v_{xy} = F(v), \qquad &&u_{xy} = F'\bigl( F^{-1}(u_x) \bigr)u, \qquad&& u=v_y;& \\
&v_{xy} = 1, \qquad &&u_{xy} = \frac{\psi''\bigl( \psi^{-1}(u) \bigr)u_y}{\psi' \bigl( \psi^{-1}(u) \bigr)}, \qquad&& u=\psi(v_y); &\\
&v_{xy} = 0, \qquad &&u_{xy} = 0, \qquad&& u = c v + \mu(v_y); &\\
&v_{xy} = 0, \qquad &&u_{xy} = - u_x \exp u, \qquad&& u = \ln v_y - \ln v; &\\
&v_{xy} = v, \qquad &&u_{xy} = u, \qquad&& u = c_1 v + c_2 v_y; &\\
&v_{xy} = 1, \qquad &&u_{xy} = 1, \qquad&& u = v + v_y,&
\end{alignat*}
up to the point transformations $u \rightarrow \theta(u)$, $v \rightarrow \kappa(v)$, $x \rightarrow \xi x$, and $y \rightarrow \eta y$, where~$\xi$ and~$\eta$ are arbitrary constants. Here $c$ is an arbitrary constant, $c_1$ and $c_2$ are constants satisfying $(c_1, c_2) \neq (0, 0)$.

Based on the above lists, B\"acklund transformations have been constructed for some pairs of equations. For instance, the equations
\begin{gather}  \label{Ex_B}
u_{xy} = F'\bigl( F^{-1}(u_x) \bigr)u,\qquad v_{xy} = F(v)
\end{gather}
are connected by the B\"acklund transformation
\[
v = F^{-1}(u_x), \qquad u = v_y.
\]
Kuznetsova~\cite{Kuzn1} showed that linearizations of equation~\eqref{Ex_B} are related by Laplace transformations of the f\/irst order. For example, we give the equations
\[
u_{xy} = \bigl( \lambda - \beta n b^{n-1}(u_x) \bigr) u , \qquad v_{xy} = \lambda v - \beta v^n, \qquad n > 0,
\]
where $\lambda$ and $\beta$ are arbitrary constants, and the function $b$ satisf\/ies the equation $\lambda b(u_x) - \beta b^n(u_x) = u_x$. The B\"acklund transformation is given by
\[
u = v_y, \qquad v = b(u_x).
\]
Note that the equation $v_{xy} = \lambda v - \beta v^n$ is a version of the PHI-four equation~\cite{SolAbdo}. The PHI-four equation and the corresponding B\"acklund transformation are obtained for $n=3$.

The purpose of this paper is to describe all equations of form \eqref{u_xy} that are transformed into equations of form \eqref{v_xy} by dif\/ferential substitutions
\begin{gather}
v = \varphi(u,u_x,u_y), \qquad \varphi_{u_x} \varphi_{u_y} \neq 0, \label{phi}
\end{gather}
over the ring of complex-valued variables.

It should be noted that most of the dif\/ferential substitutions which connect the well-known integrable equations
\eqref{u_xy} have the form $v = \varphi(u, u_x, u_y)$ (see~\cite{MeshSok, ZhSok}). Therefore, we are interested just in this form of substitutions.

This paper is organized as follows. Section~\ref{section2} presents the complete list of equations \eqref{u_xy} that are transformed into  the Klein--Gordon equations by dif\/ferential substitutions of form~\eqref{phi}. In Section~\ref{section3}, the main theorem of the paper is proven. Section~\ref{section4} is devoted to the problem which is, in a sense, inverse to the original problem. Namely, equations \eqref{v_xy} are transformed into equations \eqref{u_xy} by dif\/ferential substitutions of the form
\begin{gather}  \label{kuzn_psi}
u = \psi(v, v_y, v_x), \qquad \psi_{v_y} \psi_{v_x} \neq 0,
\end{gather}
over the ring of complex-valued variables.

\section[Equations transformed into Klein-Gordon equations]{Equations transformed into Klein--Gordon equations}\label{section2}

In this section, we give all possible cases when equation~\eqref{u_xy} is transformed into equation~\eqref{v_xy} by a dif\/ferential substitution of form~\eqref{phi}. The main result of this paper is the following theorem.
\begin{theorem}\label{theorem1}
Suppose that equation~\eqref{u_xy} is transformed into the Klein--Gordon equation \eqref{v_xy} by differential substitution \eqref{phi}. Then equations~\eqref{u_xy}, \eqref{v_xy}, and substitution \eqref{phi} take one of the following forms:
\begin{alignat}{6}
& u_{xy} = \sqrt{u^2_x + a}\sqrt{u^2_y + b}, \quad && v_{xy} =   \tfrac{1}{2}\bigl( \exp v - a b \exp(-v) \bigr),\hspace*{-200mm} &&& \nonumber\\
&&& \raisebox{0pt}[0pt][0pt]{$v = \ln \left[ \Bigl( u_x + \sqrt{u^2_x + a} \Bigr) \Bigl( u_y + \sqrt{u^2_y + b} \Bigr)\right]$};\hspace*{-200mm} &&& \label{zhiber_eq2_1}\\[0.5ex]
& u_{xy} = \sqrt{u_x u_y}, \qquad && v_{xy} = \tfrac{1}{4}v, \qquad && v = \sqrt{u_x} + \sqrt{u_y}; & \label{zhiber_eq2_2} \\
& u_{xy} = \sqrt{u_x}, \qquad && v_{xy} = \tfrac{1}{2}, \qquad && v = \sqrt{u_x} + u_y;& \label{zhiber_eq2_3}\\
& u_{xy} = 1, \qquad && v_{xy} = 0, \qquad && v = u_x + u_y; & \label{zhiber_eq2_4}
\\
& u_{xy} = \frac{1}{\gamma'(u_y)}, \qquad && v_{xy} = 1, \qquad && v = u_x + \gamma(u_y) + u,& \label{zhiber_eq2_5}
\intertext{where the function $\gamma$ satisfies $1 - \frac{\gamma''}{\gamma'^2} = \gamma'$;}
& u_{xy} = 0, \qquad && v_{xy} = 0, \qquad && v = \beta(u_x) + \gamma(u_y) + c_3 u; & \label{zhiber_eq2_6}
\\
& u_{xy} = \mu(u) u_x u_y, \qquad && v_{xy} = 0, \qquad && v = c_1 \ln u_x + c_2 \ln u_y + \alpha(u),  & \label{zhiber_eq2_7}
\intertext{where $\mu'(c_1 + c_2) + \mu^2 (c_1 + c_2) + \alpha'' + \alpha' \mu = 0$;}
& u_{xy} = \mu(u)u_x u_y, \qquad && v_{xy} = \exp v, \qquad && v = \ln (u_x u_y) + \alpha(u),& \label{zhiber_eq2_8}
\intertext{where $2 \mu' + 2 \mu^2 + \alpha'' + \alpha' \mu = \exp \alpha$;}
& u_{xy} = u, \qquad && v_{xy} = v, \qquad && v = c_1 u_y + c_2 u_x + c_3 u; & \label{zhiber_eq2_9} \\
& u_{xy} = \mu(u) (u_y + c)u_x,\qquad && v_{xy} = \exp v, \qquad && v = \ln(u_y + c) + \ln u_x + \alpha(u),\hspace*{-300mm}&  \label{zhiber_eq2_10}
\intertext{where $2 \mu' + 2\mu^2 + \alpha'' + \alpha' \mu = \exp \alpha$, $2 \mu^2 + \mu' + \alpha' \mu = \exp \alpha$;}
& u_{xy} = \mu(u)(u_y + c) u_x, \qquad && v_{xy} = 0, \qquad && v = c_2 \ln(u_y\! + c)\! + c_1 \ln u_x \!+ \alpha(u),\hspace*{-300mm} &\label{zhiber_eq2_11}
\intertext{where $(\mu' + \mu^2)(c_1 + c_2) + \alpha'' + \alpha' \mu = 0$, $c_1 \mu' + \mu^2 (c_1 + c_2) + \alpha' \mu = 0$;}
&u_{xy} = \mu(u)u_x, \qquad && v_{xy} = 0, \qquad && v = u_y - \ln u_x + \alpha(u), & \label{zhiber_eq2_12}
\intertext{where $\alpha'' + \mu' = 0$, $\mu^2 - \mu' + \alpha' \mu = 0$;}
& u_{xy} = \frac{\mu(u) u_x}{\gamma'(u_y)}, \qquad  && v_{xy} = 0, \qquad && v = \ln u_x + \gamma(u_y) + \alpha(u), &\label{zhiber_eq2_13}
\intertext{where $c_3 + \frac{\gamma''}{\gamma'^2} + c_4 \gamma' u_y = 0$, $\alpha'' + \mu' + c_4 \mu^2 = 0$, and $c_3 \mu^2 + \mu' + \mu^2 + \alpha' \mu =0$;}
& u_{xy} = \frac{u_x}{(au + b) \gamma'(u_y)}, \qquad && v_{xy} = \exp v, \qquad && v = \ln u_x + \gamma(u_y) - 2 \ln(au + b),\hspace*{-300mm} & \label{zhiber_eq2_14}
\intertext{where $c_3 + \frac{\gamma''}{\gamma'^2} + c_4 \gamma' u_y = - \gamma' \exp \gamma$, $c_3 + 1 -3a = 0$, and $c_4 + 2a^2 - a = 0$;}
& u_{xy} = - \frac{1}{u \beta'(u_x) \gamma'(u_y)}, \qquad && v_{xy} = 0, \qquad && v = \beta(u_x) + \gamma(u_y), & \label{zhiber_eq2_15}
\intertext{where $\frac{\beta''}{\beta'^2} = u_x \beta' + c_1$, $\frac{\gamma''}{\gamma'^2} = u_y \gamma' - c_1$;}
& u_{xy} = \frac{\mu(u)}{\beta'(u_x) \gamma'(u_y)}, \qquad && v_{xy} = \exp v, \qquad && v = \beta(u_x) + \gamma(u_y) + \alpha(u),\hspace*{-300mm}& \label{zhiber_eq2_16}
\intertext{where $u_x + \frac{1}{\beta'(u_x)} = \exp (\beta)$, $u_y + \frac{1}{\gamma'(u_y)} = \exp \gamma$, $\alpha'' =\exp \alpha$, and $\mu = (\exp \alpha) / \alpha'$;}
& u_{xy} = \frac{\mu(u)}{\beta'(u_x) \gamma'(u_y)}, \qquad && v_{xy} = \exp v, \qquad && v = \beta(u_x) + \gamma(u_y) + \alpha(u),\hspace*{-300mm} & \label{zhiber_eq2_17}
\intertext{where $2 u_x + \frac{1}{\beta'(u_x)} = \exp \beta$, $2 u_y + \frac{1}{\gamma'(u_y)} = \exp \gamma$, $\alpha' \mu - 2 \mu^2 = \exp \alpha$, and $\alpha'^2 = 8 \exp \alpha$;}
& u_{xy} = s(u) \sqrt{1 - u^2_x}\sqrt{1 - u^2_y}, \quad && v_{xy} = c \sin v, \qquad &&&\nonumber\\
&&&  v = \arcsin u_x + \arcsin u_y + p(u),\hspace*{-300mm} &&& \label{zhiber_eq2_18}
\intertext{where $s'' - 2 s^3 + \lambda s = 0$, $p'^2 = 2 s' - 2 s^2 + \lambda$;}
& u_{xy} = s(u) b(u_x) \bar{b}(u_y), \qquad && v_{xy} = c_1 \exp v + c_2 \exp (-2v),\hspace*{-300mm} &&& \nonumber \\
&&& \label{zhiber_eq2_19} v = -   \tfrac{1}{2} \ln(u_x - b(u_x)) - \tfrac{1}{2} \ln(u_y - \bar{b}(u_y)) + p(u), \hspace*{-50mm} &&&
\intertext{where $(u_x - b(u_x))(b(u_x) + 2u_x)^2 = 1$, $(u_y - \bar{b}(u_y))(\bar{b}(u_y) + 2 u_y)^2 = 1$, $s'' - 2 s s' - 4 s^3 = 0$, and
$p'^2 - 2 s p' - 3 s' - 2 s^2 = 0$;}
& u_{xy} = \frac{\nu(u) - q_u(u,u_y)}{q_{u_y}(u,u_y)}u_x, \qquad && v_{xy} = c_3 \exp v, \qquad && v = \ln u_x + q(u,u_y), & \label{zhiber_eq2_20}
\end{alignat}
where
\[\frac{\nu - q_u}{q_{u_y}}\left( \nu - \frac{\nu - q_u}{q^2_{u_y}} q_{u_y u_y} - 2\frac{q_{u u_y}}{q_{u_y}} \right) + \frac{\nu'}{q_{u_y}} - \frac{q_{uu}}{q_{u_y}} + \nu' u_y = c_3 \exp q, \qquad q_{uu_y} \neq 0,
\]
up to the point transformations $u \rightarrow \theta(u)$, $v \rightarrow \kappa(v)$, $x \rightarrow \xi x$, and $y \rightarrow \eta y$, and the substitution
$u + \xi x + \eta y \rightarrow u$, where $\xi$ and $\eta$ are arbitrary constants. Here $c_3$ and $c_4$ are arbitrary constants, $a$ and $b$ are constants satisfying $(a, b) \neq (0, 0)$, and $c$, $c_1$, and $c_2$ are nonzero constants; in cases~\eqref{zhiber_eq2_13} and~\eqref{zhiber_eq2_14} the function $\gamma$ satisfies the condition $\bigl( \gamma'' / \gamma'^2 \bigr)' \neq 0$; in cases~\eqref{zhiber_eq2_15}--\eqref{zhiber_eq2_17} the functions $\beta$ and $\gamma$ satisfy the conditions $\bigl( \beta'' / \beta'^2 \bigr)' \neq 0$ and $\bigl( \gamma'' / \gamma'^2 \bigr)' \neq 0$ accordingly, the function $\mu$ satisfies $\mu' \neq 0$, and $\mu \neq 0$ in all cases.
\end{theorem}

Now, let us analyze some of the above equations in detail.
Consider \eqref{zhiber_eq2_1} with $a b \neq 0$. Using the point transformations $\sqrt{a} x \rightarrow x$, $\sqrt{b} y \rightarrow y$, and $v - \ln(ab)^{1/2} \rightarrow v$, we obtain
\begin{gather} \label{zhiber_eq2_21}
u_{xy} = \sqrt{u^2_x + 1} \sqrt{u^2_y + 1}.
\end{gather}
Equation \eqref{zhiber_eq2_21} is transformed into the sine-Gordon equation
\[
v_{xy} = \frac{1}{2} \bigl( \exp v - \exp(-v) \bigr)
\]
by the dif\/ferential substitution
\[
v = \ln \Big[\Bigl( u_x + \sqrt{u^2_x + 1} \Bigr)\Bigl( u_y + \sqrt{u^2_y + 1} \Bigr)\Big].
\]
Equation \eqref{zhiber_eq2_21} is a $S$-integrable and possesses symmetries of the third order (see~\cite{MeshSok}).
Note that applying the point transformations $v \rightarrow i v$, $i x \rightarrow x$, $i y \rightarrow y$, and using the formula $\ln \big( \sqrt{1 - u^2_x} - i u_x \big) = - i \arcsin u_x$ we can also convert the above equations into
\[
u_{xy} = \sqrt{1 - u^2_x1 - u^2_y} \sqrt{1 - u^2_y}, \qquad v_{xy} = - \sin v, \qquad v = \arcsin u_x + \arcsin u_y.
\]
Now, assume that $a = 0$. Under
the transformations $v - \ln 2 \rightarrow v$, $\sqrt{b} y \rightarrow y$, and $v - \ln \sqrt{b} \rightarrow v$ equations \eqref{zhiber_eq2_1} take the form
\begin{gather} \label{zhiber_eq2_24}
u_{xy} = u_x \sqrt{u^2_y + 1},
\qquad
v_{xy} = \exp v, \qquad v = \ln u_x + \ln \Bigl( u_y + \sqrt{u^2_y + 1} \Bigr).
\end{gather}
Applying the transformation $i y \rightarrow y$ to the above equations we arrive at
\[
u_{xy} = u_x \sqrt{1 - u^2_y}, \qquad v_{xy} = -i \exp v, \qquad v = - i \arcsin u_y + \ln u_x.
\]
As shown in~\cite{MeshSok}, equation~(\ref{zhiber_eq2_24}$_1$) has symmetries of the third order. In~\cite{MeshSok} the $x$- and $y$-integrals and the general solution of equation~(\ref{zhiber_eq2_24}$_1$) were presented.

Note that the equation 
(\ref{zhiber_eq2_2}$_1$) is the Goursat equation. Its symmetries of the third order can be found, for instance, in~\cite{MeshSok}.

The equation 
(\ref{zhiber_eq2_3}$_1$) has symmetries of the third order~\cite{MeshSok}. The $x$- and $y$-integrals of this equation are given by
\[
\omega = \frac{u_{xx}}{\sqrt{u_x}}, \qquad \bar{\omega} = u_{yyy}.
\]

Consider cases \eqref{zhiber_eq2_7} and \eqref{zhiber_eq2_8}. The equation
$u_{xy} = \mu(u)u_x u_y $
possesses the $x$- and $y$-integrals of the f\/irst order, $\omega = \ln u_x - \sigma(u)$, $\bar{\omega} = \ln u_y - \sigma(u)$.
Here $\sigma' = \mu$.

The equation $
u_{xy} = \mu(u)(u_y + c)u_x
$ in cases \eqref{zhiber_eq2_10} and \eqref{zhiber_eq2_11} possess the $y$-integral of the f\/irst order $\bar{\omega} = \ln (u_y + c) - \sigma(u),$
where $\sigma' = \mu$. The $x$-integral in case \eqref{zhiber_eq2_10} is
\[
\omega = \frac{u_{xxx}}{u_x} - \frac{3}{2} \frac{u^2_{xx}}{u^2_x} - \frac{1}{2} \bigl( \mu^2(u) + 2 \mu(u) \alpha'(u) + \alpha'^2(u) \bigr)u^2_x,
\]
and in case \eqref{zhiber_eq2_11} we get the $x$-integral
\[
\omega = c_2 \mu(u) u_x + c_1 \frac{u_{xx}}{u_x} + \alpha'(u) u_x.
\]

The equation 
(\ref{zhiber_eq2_14}$_1$)
possesses the $y$-integral of the f\/irst order and the $x$-integral of the third order
\[
\bar{\omega} = \gamma(u_y) - \frac{1}{a} \ln(au + b), \qquad
\omega = \frac{u_{xxx}}{u_x} - \frac{3}{2} \frac{u^2_{xx}}{u^2_{x}} + \frac{u^2_x (2a - 1)}{2 (au + b)^2}.
\]

Now, we consider the equation
which appears in \eqref{zhiber_eq2_16} and \eqref{zhiber_eq2_17}. 
The equation~(\ref{zhiber_eq2_16}$_1$) is transformed into the equation presented in~\cite{ZhSok} by a point transformation and has the integrals of the second order
 \[
\omega = \beta'(u_x) u_{xx} - \frac{ \mu'(u)}{\mu(u) \beta'(u_x)}, \qquad \bar{\omega} = \gamma'(u_y) u_{yy} - \frac{\mu'(u)}{\mu(u) \gamma'(u_y)}.
\]
On the other hand, 
 equation~(\ref{zhiber_eq2_17}$_1$) can be transformed into the equation given in~\cite{ZhSok}
\begin{gather} \label{zhiber_eq2temp2}
u_{xy} = \frac{1}{u} B(u_x) \bar{B}(u_y).
\end{gather}
Here $B(u_x)B'(u_x) + B(u_x) - 2u_x = 0$, $\bar{B}(u_y) \bar{B}'(u_y) + \bar{B}(u_y) - 2 u_y = 0$. The integrals of equation~\eqref{zhiber_eq2temp2} are~\cite{ZhSok}
\begin{gather*}
\omega = \frac{u_{xxx}}{B} + \frac{2(B - u_x)}{B^3} u^2_{xx} + \frac{2(2u_x + B)}{u B} + \frac{B(u_x + B)}{u^2},  \\
\bar{\omega} = \frac{u_{yyy}}{\bar{B}} + \frac{2(\bar{B} - u_y)}{\bar{B}^3} u^2_{yy} + \frac{2(2u_y + \bar{B})}{u \bar{B}} + \frac{\bar{B}(u_y + \bar{B})}{u^2}.
\end{gather*}

The equation 
(\ref{zhiber_eq2_20}$_1$)
possesses the $y$-integral of the f\/irst order
$
\bar{\omega} = q(u, u_y) - \sigma(u).
$
Here $\sigma' = \nu$. If $c_3 \neq 0$ then we obtain the $x$-integral of the third order
\[
\omega = \frac{u_{xxx}}{u_x} - \frac{3}{2} \frac{u^2_{xx}}{u^2_x} + \nu'(u) u^2_x - \frac{1}{2} \nu^2(u) u^2_x.
\]
If $c_3 = 0$ then we have the $x$-integral of the second order
\[
\omega = \frac{u_{xx}}{u_x} + \nu(u) u_x.
\]

Note that equations in~\eqref{zhiber_eq2_18} and \eqref{zhiber_eq2_19} are well-known equations, which are integrable by the inverse scattering method (see~\cite{ZhSok}).

All of the previously mentioned equations possessing $x$- and $y$-integrals are contained in the list of Liouville type equations given in~\cite{ZhSok}.

Now we will show how to obtain a solution of an equation from a solution of another one by applying dif\/ferential substitutions. As an example, we consider case \eqref{zhiber_eq2_8} with specifying $\mu(u) = 1$, $\alpha(u) = \ln 2$. So we have
\[
u_{xy} = u_x u_y, \qquad v = \ln (2 u_x u_y), \qquad v_{xy} = \exp v.
\]
The equation $u_{xy} = u_x u_y$ has the $x$-integral $\omega(x) = \exp(-u)u_x$. Integrating this equation with respect to $x$ and redenoting $\int \omega(x) dx$ by $\omega(x)$ we obtain
\[
\exp(-u) = \omega(x) + \bar{\omega}(y).
\]
Hence
\[
u = -\ln \bigl(\omega(x) + \bar{\omega}(y) \bigr).
\]
Substituting the function $u$ into the equation $v = \ln(2 u_x u_y)$ we get the general solution of the Liouville equation $v_{xy} = \exp v$ as
\[
v(x,y) = \ln \left( \frac{2 \omega'(x) \bar{\omega}'(y)}{\bigl( \omega(x) + \bar{\omega}(y) \bigr)^2} \right).
\]

\section{Proof of the main theorem}\label{section3}

In this section we prove Theorem~\ref{theorem1}. In order to do that we determine the functions $f$, $F$, and $\varphi$ in \eqref{u_xy}, \eqref{v_xy} and \eqref{phi}. By substituting function \eqref{phi} into equation~\eqref{v_xy} and using equation~\eqref{u_xy} we get
\begin{gather}
 \varphi_u f + u_x \bigl( \varphi_{uu} u_y + \varphi_{u u_x} f + \varphi_{u u_y} u_{yy} \bigr) + u_{xx} \bigl( \varphi_{u_x u} u_y + \varphi_{u_x u_x} f + \varphi_{u_x u_y} u_{yy} \bigr) \nonumber\\
\qquad{} + \varphi_{u_x}\bigl( f_u u_x + f_{u_x} u_{xx} + f_{u_y} f\bigr) + \varphi_{u_y} \bigl( f_u u_y + f_{u_x} f + f_{u_y} u_{yy} \bigr) \nonumber\\
\qquad{}+ f \left( \varphi_{u_y u} u_y + \varphi_{u_y u_x} f + \varphi_{u_y u_y} u_{yy} \right) = F(\varphi). \label{zhiber4}
\end{gather}
Since the function $F(\varphi)$ depends only on $u$, $u_x$, and $u_y$,
the coef\/f\/icients at $u_{xx}$, $u_{yy}$, and $u_{xx}u_{yy}$ are equal to zero, i.e.
\begin{gather*}
\varphi_{u_x u_y} = 0, \qquad
\varphi_{u u_x} u_y + \varphi_{u_x u_x} f + \varphi_{u_x} f_{u_x} = 0,\qquad
\varphi_{u u_y} u_x + \varphi_{u_y} f_{u_y} + f \varphi_{u_y u_y} = 0.
\end{gather*}
Integration of these equations leads to
\begin{gather}
\varphi = p(u, u_x) + q(u, u_y),   \label{firstcondition}\\
\varphi_u u_y + \varphi_{u_x} f = A(u, u_y), \label{secondcondition}\\
\varphi_u u_x + \varphi_{u_y} f = B(u, u_x). \label{thirdcondition}
\end{gather}
The remaining terms in \eqref{zhiber4} give 
\begin{gather} \label{fourthcondition}
f \bigl( \varphi_u\! + u_x \varphi_{u u_x} \!+ \varphi_{u_x} f_{u_y}\! + \varphi_{u_y} f_{u_x}\! + u_y \varphi_{u u_y} \bigr)\! + \varphi_{u u} u_x u_y\! + \big( u_x \varphi_{u_x}\! + u_y \varphi_{u_y} \big) f_u = F(\varphi).\!\!\!\!
\end{gather}
Hence, the original classif\/ication problem is reduced to the analysis of equations~\eqref{firstcondition}--\eqref{fourthcondition}. Eliminating the function $f$ from equations~\eqref{secondcondition} and \eqref{thirdcondition} we obtain the relation
\begin{gather}
\bigl( A - u_y \varphi_u \bigr) \varphi_{u_y} = \bigl( B - u_x \varphi_u \bigr) \varphi_{u_x}. \label{zhiber12}
\end{gather}
Applying the operator $\frac{\partial^2}{\partial u_x \partial u_y}$ to equation~\eqref{zhiber12} we arrive at the equation
\begin{gather}
\bigl( u_y \varphi_{u_y} \bigr)_{u_y} \varphi_{u u_x} = \bigl( u_x \varphi_{u_x} \bigr)_{u_x} \varphi_{u u_y}. \label{zhiber13}
\end{gather}
Relation \eqref{zhiber13} is satisf\/ied if one of the following conditions hold:
\begin{gather}
\varphi_{u u_x} = 0, \qquad \varphi_{u u_y} = 0,  \label{zhiber14}\\
\varphi_{u u_x} = 0, \qquad \bigl( u_x \varphi_{u_x} \bigr)_{u_x} = 0, \label{zhiber15}\\
( u_y \varphi_{u_y} )_{u_y} = 0, \qquad \varphi_{u u_y} = 0, \label{zhiber16}\\
( u_y \varphi_{u_y} )_{u_y} = 0, \qquad ( u_x \varphi_{u_x} )_{u_x} = 0, \label{zhiber17}\\
\frac{( u_y \varphi_{u_y} )_{u_y}}{\varphi_{u u_y}} = \frac{( u_x \varphi_{u_x} )_{u_x}}{\varphi_{u u_x}} = \lambda(u), \qquad \lambda(u) \neq 0. \label{zhiber18}
\end{gather}

First, let us analyze equation~\eqref{zhiber18}. By substituting the function $\varphi$ given by \eqref{firstcondition} into equation~\eqref{zhiber18} we get
\begin{gather}\label{zhiber18.5}
( u_y q_{u_y} )_{u_y} = \lambda(u) q_{u u_y}, \qquad ( u_x p_{u_x} )_{u_x} = \lambda(u) p_{u u_x}.
\end{gather}
Now we integrate the f\/irst equation of \eqref{zhiber18.5} with respect to $u_y$ and the second one with respect to $u_x$. This gives
\[
u_y q_{u_y} = \lambda(u) q_u + C(u), \qquad u_x p_{u_x} = \lambda(u) p_u + E(u).
\]
The general solutions of these equations are
\[
q = \Phi_1 ( u_y \kappa(u)  ) + \epsilon(u), \qquad p = \Phi_2  ( u_x \kappa(u)  ) + \mu(u),
\]
where $
\kappa(u) = \lambda(u) \kappa'(u)$, $\lambda(u) \epsilon'(u) + C(u) = 0$, $\lambda(u) \mu'(u) + E(u) = 0$. Therefore, the function $\varphi$ def\/ined by \eqref{firstcondition} takes the form 
\[
\varphi = \Phi(u) + \Phi_1  ( u_y \kappa(u)  )+ \Phi_2  ( u_x \kappa(u)  ).
\]
Here $\Phi(u) = \epsilon(u) + \mu(u)$. Furthermore, if we use the point transformation
$
\int{\kappa(u) du} \rightarrow u
$
in the above formula, we obtain
\begin{gather}
\varphi = \alpha(u) + \beta(u_x) + \gamma(u_y). \label{zhiber19}
\end{gather}
Clearly, function \eqref{firstcondition} satisfying \eqref{zhiber14} also takes form \eqref{zhiber19}.

Assume that condition \eqref{zhiber15} holds. In this case, the substitution of the functions $\varphi$ def\/ined by \eqref{firstcondition} into  \eqref{zhiber15} yields
\[
p_{u u_x} = 0, \qquad  ( u_x p_{u_x}  )_{u_x} = 0,
\]
which gives
\[
p = \alpha(u) + c \ln u_x.
\]
Here $c$ is an arbitrary constant. Hence, function \eqref{firstcondition} takes the form $
\varphi = \alpha(u) + c \ln u_x + q(u, u_y)
$.
Replacing $\alpha(u) + q(u, u_y)$ by $q(u, u_y)$ in this equation we get
\begin{gather}
\varphi = c \ln u_x + q(u, u_y). \label{zhiber20}
\end{gather}
Recall that 
$\varphi_{u_x} \varphi_{u_y} \neq 0$. This property implies $c \neq 0$.
Clearly, case \eqref{zhiber16} coincides with \eqref{zhiber15} up to the permutation of~$x$ and~$y$.

It remains to consider the case when $\varphi$ satisf\/ies \eqref{zhiber17}. Based on \eqref{firstcondition}, we rewrite \eqref{zhiber17} as
\[
( u_y q_{u_y} )_{u_y} = 0, \qquad ( u_x p_{u_x} )_{u_x} = 0.
\]
By integrating these equations we get the functions $q$ and $p$,
\[
q = \mu(u) \ln u_y + \epsilon(u), \qquad p = \kappa(u) \ln u_x + \delta(u).
\]
Consequently, the function $\varphi$ def\/ined by formula \eqref{firstcondition} takes the form
\begin{gather}
\varphi = \alpha(u) + \kappa(u) \ln u_x + \mu(u) \ln u_y. \label{zhiber21}
\end{gather}
Thus, to solve the original classif\/ication problem it is suf\/f\/icient to consider three cases: \eqref{zhiber19}, \eqref{zhiber20}, and \eqref{zhiber21}.

\subsection[Case $\varphi = \alpha(u) + \beta(u_x) + \gamma(u_y)$]{Case $\boldsymbol{\varphi = \alpha(u) + \beta(u_x) + \gamma(u_y)}$}

When we substitute \eqref{zhiber19} into equation \eqref{zhiber12}, we obtain
\[
\bigl( A(u, u_y) - u_y \alpha'(u) \bigr) \gamma'(u_y) = \bigl( B(u, u_x) - u_x \alpha'(u) \bigr) \beta'(u_x).
\]
Since $u_x$ and $u_y$ are regarded as independent variables, 
the above equation is equivalent to the system
\begin{gather*}
\bigl( A(u, u_y) - u_y \alpha'(u) \bigr) \gamma'(u_y) = \mu(u),\qquad
\bigl( B(u, u_x) - u_x \alpha'(u) \bigr) \beta'(u_x)= \mu(u).
\end{gather*}
From this system we f\/ind the functions $A$ and $B$ as
\[
A = \frac{\mu}{\gamma'} + u_y \alpha', \qquad B = \frac{\mu}{\beta'} + u_x \alpha'.
\]
By substituting $A$ and $B$ into equations~\eqref{secondcondition} and \eqref{thirdcondition} we determine $f$ as follows
\begin{gather}
f = \frac{\mu(u)}{\beta'(u_x) \gamma'(u_y)}. \label{zhiber22}
\end{gather}
Using \eqref{zhiber22} we transform equation~\eqref{fourthcondition} into
\begin{gather}
\frac{\alpha' \mu}{\beta' \gamma'} - \mu^2 \left( \frac{\gamma''}{\gamma'^2} + \frac{\beta''}{\beta'^2} \right) \frac{1}{\beta' \gamma'} + \alpha'' u_x u_y + \mu' \left( \frac{u_x}{\gamma'} + \frac{u_y}{\beta'} \right) = F(\alpha + \beta + \gamma).    \label{zhiber23}
\end{gather}
Applying the operators $\frac{\partial}{\partial u_x}$ and $\frac{\partial}{ \partial u_y}$ to equation~\eqref{zhiber23} we obtain 
\begin{gather*}
-\alpha' \mu \frac{\beta''}{\beta'^2 \gamma'} -\mu^2 \left( -\frac{\gamma''}{\gamma'^3} \frac{\beta''}{\beta'^2} + \frac{1}{\gamma'} \left( \frac{\beta''}{\beta'^3} \right)' \right) + \alpha'' u_y + \mu' \left( \frac{1}{\gamma'} - u_y \frac{\beta''}{\beta'^2} \right) = F'(\alpha + \beta + \gamma) \beta',\\
-\alpha' \mu \frac{\gamma''}{\beta' \gamma'^2} - \mu^2 \left( \left( \frac{\gamma''}{\gamma'^3} \right)' \frac{1}{\beta'} - \frac{\beta''}{\beta'^3} \frac{\gamma''}{\gamma'^2} \right) + \alpha'' u_x + \mu' \left( -u_x \frac{\gamma''}{\gamma'^2} + \frac{1}{\beta'} \right) = F'(\alpha + \beta + \gamma) \gamma'.
\end{gather*}
By eliminating $F'$ from these equations we get
\begin{gather}
-\alpha' \mu \frac{\beta''}{\beta'^2} - \mu^2 \left( \frac{\beta''}{\beta'^3} \right)' + \alpha'' u_y \gamma' - \mu' u_y \gamma' \frac{\beta''}{\beta'^2}
\nonumber\\
\qquad{} = -\alpha' \mu \frac{\gamma''}{\gamma'^2} - \mu^2 \left( \frac{\gamma''}{\gamma'^3} \right)' + \alpha'' u_x \beta' - \mu' u_x \beta' \frac{\gamma''}{\gamma'^2}.
\label{zhiber24}
\end{gather}
Under the action of the operator $\frac{\partial^2}{\partial u_x \partial u_y}$, equation~\eqref{zhiber24} takes the form
\[
\mu' \left( (u_x \beta')' \left( \frac{\gamma''}{\gamma'^2} \right)' - (u_y \gamma')' \left( \frac{\beta''}{\beta'^2} \right)' \right) = 0.  \]
It can be easily seen that the above equation is true if one of the following conditions is met:
\begin{gather}
\mu'(u) = 0,  \label{zhiber26}\\
(u_x \beta')' = 0, \qquad (u_y \gamma')' = 0, \label{zhiber27}\\
(u_x \beta')' = 0, \qquad \left( \frac{\beta''}{\beta'^2} \right)' = 0, \label{zhiber28}\\
\left( \frac{\gamma''}{\gamma'^2} \right) ' = 0, \qquad (u_y \gamma')' = 0, \label{zhiber29}\\
\left( \frac{\gamma''}{\gamma'^2} \right)' = 0, \qquad \left( \frac{\beta''}{\beta'^2} \right)' = 0, \label{zhiber30}\\
 \frac{(u_x\beta')' }{\left( \displaystyle \frac{\beta''}{\beta'^2} \right)'} = \frac{ (u_y\gamma')' }{ \left( \displaystyle \frac{ \gamma''}{\gamma'^2} \right)' } \neq 0. \label{zhiber31}
\end{gather}
It should be noted that $\mu' \neq 0$ in cases \eqref{zhiber27}--\eqref{zhiber31}.

To analyze cases \eqref{zhiber26}--\eqref{zhiber31} in a unif\/ied manner we begin by giving the following lemma. 
\begin{lemma}\label{lemma1}
By condition \eqref{zhiber26}, equations~\eqref{u_xy}, \eqref{v_xy}, and substitution \eqref{phi} take one of the following forms:
\begin{alignat}{6}
& u_{xy} = 0, \qquad && v_{xy} = \exp v, \qquad && v = \alpha(u) + \ln (u_x u_y), & \label{zhiber32}
\intertext{where the function $\alpha$ satisfies $\alpha'' = \exp \alpha$;}
& u_{xy} = u_x u_y, \qquad && v_{xy} = \exp v, \qquad && v = \alpha(u) + \ln (u_x u_y),& \label{zhiber33}
\intertext{where $\alpha'' + \alpha' + 2 = \exp \alpha$;}
& u_{xy} = - u_x u_y, \qquad && v_{xy} = 0, \qquad && &\nonumber\\
&&& v = \exp u + (a_1 + b_1)u + a_1 \ln u_x + b_1 \ln u_y;\hspace*{-300mm} &&&  \label{zhiber34}\\
& u_{xy} = c \sqrt{u^2_x + a_2} \sqrt{u^2_y + b_2}, \qquad &&  v_{xy} = \frac{c^2}{2} \bigl( \exp v - a_2 b_2 \exp(-v) \bigr),\hspace*{-300mm} &&& \nonumber \\
&&& v = \ln \left[ \left( u_x + \sqrt{u^2_x + a_2} \right) \left( u_y + \sqrt{u^2_y + b_2} \right) \right];\hspace*{-300mm} &&& \label{zhiber35}
\\
& u_{xy} = c \sqrt{1 - u^2_x} \sqrt{1 - u^2_y}, \qquad && v_{xy} = -c^2 \sin v, \qquad && v = \arcsin u_x + \arcsin u_y;\hspace*{-100mm} & \label{zhiber36}\\
& u_{xy} = c \sqrt{u_xu_y}, \qquad && v_{xy} = \frac{c^2 v}{4}, \qquad && v = \sqrt{u_x} + \sqrt{u_y};& \label{zhiber37}\\
& u_{xy} = c \sqrt{u_x}, \qquad && v_{xy} = \frac{c^2}{2}, \qquad && v = \sqrt{u_x} + u_y; & \label{zhiber38}\\
& u_{xy} = c, \qquad && v_{xy} = 0, \qquad && v = u_x + u_y;& \label{zhiber39}\\
& u_{xy} = c u_y \sqrt{1 - u^2_x}, \qquad && v_{xy} =- \mathrm{i} c^2 \exp v, \qquad && v = -\mathrm{i} \arcsin u_x + \ln u_y;\hspace*{-100mm} & \label{zhiber40}\\
& u_{xy} = \frac{a_1}{\gamma'(u_y)}, \qquad && v_{xy} = b_1, \qquad && v = u_x + \gamma(u_y) + u,&  \label{zhiber41}
\intertext{where $a_1 - a^2_1 \frac{\gamma''}{\gamma'^2} = b_1 \gamma'$;}
& u_{xy} = a(u_x + c_7)(u_y + c_9), \qquad && v_{xy} = 0, \qquad && \nonumber\\
&&& v = a_1 \ln(u_x + c_7) + b_1 \ln (u_y + c_9) + u,\hspace*{-300mm} &&& \label{zhiber42}
\intertext{where $aa_1 + ab_1 + 1 = 0$;}
& u_{xy} = 0, \qquad &&  v_{xy} = 0, \qquad  && v = \beta(u_x) + \gamma(u_y) + u, & \label{zhiber43}
\end{alignat}
up to the point transformations $u \rightarrow \theta(u)$, $v \rightarrow \kappa(v)$, $x \rightarrow \xi x$, and $y \rightarrow \eta y$ and the substitution 
$u + \xi x + \eta y \rightarrow u$, where $\xi$ and $\eta$ are arbitrary constants. Here $\alpha''$, $\alpha'$, and $1$ are linearly independent functions, $c$, $c_1$, $c_2$, $c_7$, $c_9$, $a_1 \neq 0$, $b_1 \neq 0$, $a \neq 0$, $b_2$, and $a_2$ are arbitrary constants.
\end{lemma}

\begin{proof} If condition \eqref{zhiber26} holds then $\mu(u) = c$, where $c$ is an arbitrary constant. Re\-wri\-ting~\eqref{zhiber24} we obtain
\begin{gather*}
c \alpha'(u) \frac{\beta''(u_x)}{\beta'^2(u_x)} + c^2 \left( \frac{\beta''(u_x)}{\beta'^3(u_x)} \right)' + \alpha''(u)  u_x \beta'(u_x) \\
\qquad{} = c \alpha'(u) \frac{\gamma''(u_y)}{\gamma'^2(u_y)} + c^2 \left( \frac{\gamma''(u_y)}{\gamma'^3(u_y)} \right)' + \alpha''(u) u_y \gamma'(u_y).
\end{gather*}
Since we regard the variables $u_x$, $u_y$ as independent, this equation is equivalent to the equations
\begin{gather*}
 c \alpha'(u) \frac{\beta''(u_x)}{\beta'^2(u_x)} + c^2 \left( \frac{\beta''(u_x)}{\beta'^3(u_x)} \right)' + \alpha''(u) u_x \beta'(u_x) = \sigma(u),\\
 c \alpha'(u) \frac{\gamma''(u_y)}{\gamma'^2(u_y)} + c^2 \left( \frac{\gamma''(u_y)}{\gamma'^3(u_y)} \right)' + \alpha''(u) u_y \gamma'(u_y) = \sigma(u).
\end{gather*}
By the same fact that the variables $u_x$, $u_y$ are considered as independent we def\/ine the func\-tion~$\sigma$ as
$
\sigma(u) = A_1 \alpha'(u) + B_1 \alpha''(u) + C_1 
$.
 According to this we rewrite the above equations as
\begin{gather}
\alpha' \left( c  \frac{\beta''}{\beta'^2} - A_1 \right) + \alpha'' \left( u_x \beta' - B_1\right) =  C_1 - c^2 \left( \frac{\beta''}{\beta'^3} \right)', \nonumber\\
 \alpha' \left( c  \frac{\gamma''}{\gamma'^2} - A_1 \right) + \alpha'' \left( u_y \gamma' - B_1 \right) =  C_1 - c^2 \left( \frac{\gamma''}{\gamma'^3} \right)'.
 \label{zhiber46}
\end{gather}
Here $A_1$, $B_1$, and $C_1$ are constants.

Let us assume that $1$, $\alpha'$, and $\alpha''$ are linearly independent functions. Clearly, equations~\eqref{zhiber46} imply
\begin{gather*}
c \frac{\beta''}{\beta'^2} = A_1, \qquad u_x \beta' = B_1, \qquad C_1 - c^2 \left( \frac{\beta''}{\beta'^3} \right)' = 0, \\
c \frac{\gamma''}{\gamma'^2} = A_1, \qquad u_y \gamma' = B_1, \qquad C_1 - c^2 \left( \frac{\gamma''}{\gamma'^3} \right)' = 0.
\end{gather*}
From these equations we get
\[
\beta' = \frac{B_1}{u_x}, \qquad \gamma' = \frac{B_1}{u_y}, \qquad -\frac{c}{B_1} = A_1, \qquad C_1 + \frac{c^2}{B^2_1} = 0.
\]
Using the above equations we transform equation~\eqref{zhiber23} into the equation
\begin{gather}
u_x u_y \left( \frac{c \alpha'}{B^2_1} + \frac{2 c^2}{B^3_1} + \alpha'' \right) = F(\alpha + \beta + \gamma). \label{zhiber47}
\end{gather}
Since $1$, $\alpha'$, and $\alpha''$ are linearly independent functions, the left-hand side of equation~\eqref{zhiber47} does not vanish. Then $F \neq 0$. By dif\/ferentiating \eqref{zhiber47} with respect to $u_x$ and using $\beta' = B_1 / u_x$ we get the equation $1 = F'(z)B_1/F(z)$, where $z = \alpha + \beta + \gamma$. Its general solution is given by
\begin{gather}
F(z) = C_1 \exp(z/B_1). \label{zhiber48}
\end{gather}
Substituting function \eqref{zhiber48} into equation~\eqref{zhiber47} and using $\beta = B_1\ln u_x + C_2$, $\gamma = B_1 \ln u_y + C_3$ we obtain
\[
\frac{c \alpha'}{B^2_1} + \frac{2c^2}{B^3_1} + \alpha'' = C_1 \exp\left( \frac{\alpha}{B_1} + \frac{C_2}{B_1} + \frac{C_3}{B_1} \right).
\]
Thus, equations~\eqref{u_xy}, \eqref{v_xy}, and \eqref{phi} have the following forms
\[
{\displaystyle u_{xy} = \frac{c u_x u_y}{B^2_1}, \qquad v_{xy} = C_1 \exp(v/B_1), \qquad v = \alpha(u) + B_1 \ln( u_x u_y) + C_2 + C_3,}
\]
where
\[
{\displaystyle c \frac{\alpha'}{B^2_1} + 2c^2 \frac{1}{B^3_1} + \alpha'' = C_1 \exp\left( \frac{\alpha + C_2 + C_3}{B_1} \right).}
\]
We redenote $(\alpha + C_2 + C_3) / B_1$ by $\alpha$. Under 
the point transformation $v \rightarrow B_1 v$ the above equations take the forms 
\[
{\displaystyle u_{xy} = \frac{c u_x u_y}{B^2_1}, \qquad v_{xy} = \frac{C_1}{B_1} \exp v, \qquad v = \alpha(u) + \ln (u_x u_y),}
\]
where
\[
{\displaystyle c \frac{\alpha'}{B^2_1} + 2 c^2 \frac{1}{B^4_1} + \alpha'' = \frac{C_1}{B_1} \exp \alpha.}
\]
The multiplier $C_1/B_1$ can be eliminated by the shift $v \rightarrow v + \ln(B_1/C_1)$. Finally, redenoting $\alpha - \ln(B_1/C_1)$ by $\alpha $ and $c/B^2_1$ by $c$ we get
\[
u_{xy} = c u_x u_y, \qquad v_{xy} = \exp v, \qquad v = \alpha(u) + \ln (u_x u_y),
\]
where $\alpha'' + c\alpha' + 2c^2 = \exp \alpha$.
If $c = 0$ then these equations take the form 
\eqref{zhiber32}. Otherwise, applying the point transformation $u \rightarrow u/c$ and redenoting $\alpha$ by $\alpha+\ln c^2$ we can reduce the above equations to form \eqref{zhiber33}.

Let us assume that $1$, $\alpha'$, and $\alpha''$ are linearly dependent functions. It means that
\[
C_1 \alpha'' + C_2 \alpha' + C_3 = 0, \qquad (C_1, C_2, C_3) \neq (0, 0, 0).
\]
If $C_1 = 0$ then $C_2 \neq 0$ and we get
$
\alpha' = c.
$
Otherwise,
$
\alpha'' = c_1 \alpha' + c_2.  \label{zhiber53}
$
Case $\alpha' = c$ is a subcase of $\alpha'' = c_1 \alpha' + c_2$. This equation has two families of solutions
\[
\alpha = c_3 u^2 + c_4 u + c_5, \qquad \alpha = \frac{1}{c_1} \exp (c_1u) + c_6 u + c_7.
\]
The constants $c_5$, $c_7$ can be eliminated by $\beta + c_5 \rightarrow \beta$, $\beta + c_7 \rightarrow \beta$ in equation~\eqref{zhiber19}. So there are two possibilities
\begin{gather}
\alpha = c_2 u^2 + c_3 u \label{zhiber54}
\end{gather}
and
\[
\alpha = \bigl( \exp c_1 u \bigr)/ c_1 + c_4 u,
\]
which takes the form 
\begin{gather}
\alpha = \exp (c_1u) + c_4 u, \qquad c_1 \neq 0 \label{zhiber55}
\end{gather}
under the shifts $u \rightarrow u + (\ln c_1) / c_1 $ and $\alpha \rightarrow \alpha + c_4 (\ln c_1) / c_1$.

Now, let us concentrate on case \eqref{zhiber55}, taking into account the fact that $\mu(u) = c$. Equation~\eqref{zhiber23} can be rewritten as
\begin{gather}
\frac{c (c_1 \exp (c_1u) + c_4)}{\beta' \gamma'} - c^2 \left( \frac{\gamma''}{\gamma'^2} + \frac{\beta''}{\beta'^2} \right) \frac{1}{\beta' \gamma'} + c^2_1 \exp (c_1 u) u_x u_y = F(\alpha + \beta + \gamma). \label{zhiber56}
\end{gather}
Applying $\frac{\partial}{\partial u}$ to equation~\eqref{zhiber56} we obtain
\[
\frac{c c^2_1 \exp (c_1 u)}{\beta' \gamma'} + c^3_1 \exp( c_1 u) u_x u_y = F'(\alpha + \beta + \gamma) (c_1 \exp (c_1u) + c_4).
\]
Therefore,
\[
\frac{c c^2_1}{\beta' \gamma'} + c^3_1 u_x u_y = (c_1 + c_4 \exp(-c_1u)) F'(\alpha + \beta + \gamma).
\]
Next, by applying the dif\/ferentiation $\frac{\partial}{\partial u}$ to both sides of this equation, we get
\[
-c_1 c_4 \exp(-c_1 u) F'(\alpha + \beta + \gamma) + (c_1 + c_4 \exp(-c_1 u)) (c_1 \exp (c_1 u) + c_4) F''(\alpha + \beta + \gamma) = 0.
\]
It is not dif\/f\/icult to see that the above equation implies
\[
(c_1 \exp (c_1 u) + c_4)^2 F''(\alpha + \beta + \gamma) = c_1 c_4 F'(\alpha + \beta + \gamma).
\]
Consequently, we have two possibilities
\begin{gather}
F'(\alpha + \beta + \gamma) = 0, \label{zhiber57}\\
\frac{F''(\alpha + \beta + \gamma)}{F'(\alpha + \beta + \gamma)} = \frac{c_1 c_4}{(c_1 \exp (c_1 u) + c_4)^2}. \label{zhiber58}
\end{gather}

Equation \eqref{zhiber57} yields $F = c_5$, where $c_5$ is an arbitrary constant. In this case by using \eqref{zhiber56} we obtain
\[
\frac{c c_1}{\beta' \gamma'} + c^2_1 u_x u_y = 0, \qquad \frac{c c_4}{\beta' \gamma'} - c^2 \left( \frac{\gamma''}{\gamma'^2} + \frac{\beta''}{\beta'^2} \right) \frac{1}{\beta' \gamma'} = c_5.
\]
According to the fact that
$u_x$ and $u_y$ are considered as independent variables we have
\[
\beta'(u_x) = \frac{c c_1}{c_6 u_x}, \qquad \gamma'(u_y) = -\frac{c_6}{c^2_1 u_y}, \qquad cc_4 - c^2 \left( \frac{c^2_1}{c_6} - \frac{c_6}{c_1c} \right) = c_5 \beta'(u_x) \gamma'(u_y).
\]
Moreover, since
$\beta' \gamma' \neq 0$ we get $c_5 = 0$, hence $F \equiv 0$. Consequently, equations~\eqref{u_xy}, \eqref{v_xy}, and~\eqref{phi} take the following forms
\[
 u_{xy} = -c_1 u_x u_y, \qquad v_{xy} = 0, \qquad v = \exp c_1u + c_4 u + \frac{cc_1}{c_6} \ln u_x - \frac{c_6}{c^2_1} \ln u_y + c_7,
\]
where
\[
 c_4 - \frac{cc^2_1}{c_6} + \frac{c_6}{c_1} = 0, \qquad c_1 \neq 0.
\]
Using the point transformations $ u \rightarrow u / c_1$, $v \rightarrow v - cc_1 \ln (c_1) / c_6 + c_6 \ln( c_1) / c^2_1 + c_7$, and redenoting $cc_1/c_6$ by $a_1$, $-c_6 / c^2_1$ by $b_1$ we get equation~\eqref{zhiber34}.

Now, suppose that \eqref{zhiber58} is true. Applying $\frac{\partial}{\partial u_x}$ to both sides of equation~\eqref{zhiber58} we get
\[
\left( \frac{F''}{F'} \right)' \beta' = 0.
\]
Recall that $\beta' \neq 0$, therefore $F'' / F' = 0$. This equation has two families of solutions. Namely, $F(z) = c_6 \exp c_5 z + c_7$, $c_5 c_6 \neq 0,
$
which turns into
\begin{gather}
F(z) = \exp c_5 z + c_7, \qquad c_5 \neq 0 \label{zhiber60}
\end{gather}
by the shift $z \rightarrow z - \left( \ln c_6 \right)/c_5$, and
\begin{gather}
F(z) = c_6 z + c_7, \qquad c_6 \neq 0. \label{zhiber61}
\end{gather}

Now consider equation~\eqref{zhiber60}. In this case, equation~\eqref{zhiber56} takes the form 
\begin{gather*}
 \frac{c(c_1 \exp (c_1 u) + c_4)}{\beta' \gamma'} - c^2 \left( \frac{\gamma''}{\gamma'^2} + \frac{\beta''}{\beta'^2} \right) \frac{1}{\beta' \gamma'} + c^2_1 \exp (c_1u) u_x u_y \\
\qquad{} = \exp \bigl( c_5(\exp c_1u + c_4 u) \bigr) \exp( c_5 \beta ) \exp (c_5 \gamma).
\end{gather*}
This equation is not satisf\/ied because $c_5 c_1 \neq 0$.

Let us focus on equation~\eqref{zhiber61}. Equation \eqref{zhiber56} can be written as
\begin{gather*}
\frac{c (c_1 \exp (c_1 u) + c_4)}{\beta' \gamma'} - c^2 \left( \frac{\gamma''}{\gamma'^2} + \frac{\beta''}{\beta'^2} \right) \frac{1}{\beta' \gamma'} + c^2_1 \exp(c_1 u ) u_x u_y \\
\qquad{} = c_6 (\exp (c_1u) + c_4 u + \beta + \gamma) + c_7.
\end{gather*}
Applying the operator $\frac{\partial}{\partial u}$ to the above equation gives 
\[
\frac{c c^2_1 \exp (c_1 u)}{\beta' \gamma'} + c^3_1 u_x u_y \exp (c_1 u) = c_6 (c_1 \exp c_1 u + c_4).
\]
Collecting the coef\/f\/icients at $\exp(c_1 u)$ and rewriting the remaining terms we obtain
\[
\frac{cc^2_1}{\beta' \gamma'} + c^3_1 u_x u_y = c_6 c_1, \qquad c_6 c_4 = 0.
\]
Since $u_x$ and $u_y$ are considered as independent, the f\/irst equation is true if and only if $c_6 = 0$. In this case, it is clear that we obtain the equations 
\eqref{zhiber34}.

Assume that the function $\alpha$ satisf\/ies equation~\eqref{zhiber54}. Using \eqref{zhiber54} and $\mu(u) = c$ we transform equation~\eqref{zhiber23} into
\[
\frac{c}{\beta' \gamma'} (2c_2u + c_3) - c^2 \left( \frac{\gamma''}{\gamma'^2} + \frac{\beta''}{\beta'^2} \right) \frac{1}{\beta' \gamma'} + 2c_2 u_x u_y = F(c_2 u^2 + c_3 u + \beta + \gamma).
\]
Dif\/ferentiating this equation with respect to $u$ and denoting $c_2 u^2 + c_3 u + \beta + \gamma$ by $z$ we obtain
\begin{gather}
2c_2 \frac{c}{\beta' \gamma'} = F'(z) (2c_2u + c_3). \label{zhiber62}
\end{gather}

Now we should analyze equation~\eqref{zhiber62}. First, we suppose that $c_2 = c_3 = 0$. The function $\alpha$ described by equation \eqref{zhiber54} vanishes. Equations \eqref{zhiber46} can be written as
\[
c_1 u_x - c^2 \frac{\beta''}{\beta'^3} = a_1, \qquad c_1 u_y - c^2 \frac{\gamma''}{\gamma'^3} = b_1.
\]
Here $a_1$, $b_1$ are arbitrary constants. The above equations imply
\begin{gather*}
\beta'(u_x) = \sqrt{-c^2} \frac{1}{\sqrt{c_1 u^2_x - 2a_1u_x + 2 a_2}}, \qquad
\gamma'(u_y) = \sqrt{-c^2} \frac{1}{\sqrt{c_1 u^2_y - 2b_1 u_y + 2 b_2 }}.
\end{gather*}
Integrating these equations we obtain distinct formulae which determine the functions $\beta$ and $\gamma$. Uniting these formulae in pairs we arrive at
\eqref{zhiber35}--\eqref{zhiber40}.

Furthermore, we must consider equation~\eqref{zhiber62} if $c_2 \neq 0$, $c_3 = 0$, and $c_2 c_3 \neq 0$.
Taking the logarithm of both sides of equation~\eqref{zhiber62} leads to
\[
\ln \left( 2c_2 \frac{c}{\beta' \gamma'} \right) = \ln F'(z) + \ln (2c_2 u + c_3).
\]
To eliminate $\beta'(u_x)$ and $\gamma'(u_y)$ we dif\/ferentiate this equation with respect to $u$,
\begin{gather}
0 = \frac{F''}{F'} (2c_2u + c_3) + \frac{2c_2}{2c_2u+c_3}. \label{zhiber77}
\end{gather}
Applying $\frac{\partial}{\partial u_x}$ to both sides of equation~\eqref{zhiber77} we get $(F''/ F')' = 0$, which means that
$F''/F = c_4$. By virtue of this, equation~\eqref{zhiber77} is written as{\samepage
\[
c_4 (2 c_2u + c_3)^2 + 2c_2 = 0.
\]
Hence $c_2 = 0$. This contradicts $c_2 \neq 0$.}

It remains to discuss the case if $c_2 = 0$, $c_3 \neq 0$.
It is clear that we have $F(z) = c_4$ from equation~\eqref{zhiber62}. Here $c_4$ is an arbitrary constant. Rewriting \eqref{zhiber23} with $\alpha = c_3 u$, $\mu = c$ we get
\begin{gather}
c_3 c - c^2 \left( \frac{\gamma''}{\gamma'^2} + \frac{\beta''}{\beta'^2} \right) = c_4 \beta' \gamma'. \label{zhiber78}
\end{gather}
The equation
\[
-c^2 \left( \frac{\beta''}{\beta'^2} \right)' = c_4 \beta'' \gamma',
\]
arises when we apply $\frac{\partial}{\partial u_x}$ to both the sides of equation~\eqref{zhiber78}.

Suppose that $\beta'' = 0$. Determining the function $\beta$ as $\beta(u_x) = c_5 u_x + c_6$, we transform equation~\eqref{zhiber78} into an ordinary dif\/ferential equation
\[
c_3 c - c^2 \frac{\gamma''}{\gamma'^2} = c_4 c_5 \gamma'.
\]
 Thus, we f\/ind equations of forms~\eqref{u_xy}, \eqref{v_xy}, and \eqref{phi},
\[
u_{xy} = \frac{c}{c_5 \gamma'(u_y)}, \qquad v_{xy} = c_4, \qquad v = c_5 u_x + \gamma(u_y) + c_3 u,
\]
where $c_3 c - c^2 \gamma''/ \gamma'^2 = c_4 c_5 \gamma'$, $c_5 \neq 0$.
We use the transformations $x / c_5 \rightarrow x$, $v / c_3 \rightarrow v$. Then we redenote $c_4 c_5 $ by $c_2$, $\gamma / c_3$ by $\gamma$. To obtain \eqref{zhiber41} we apply the transformation $c_3 x \rightarrow x$ once again. Finally, we redenote $c / c^2_3$ by $a_1$, $c_2 / c^2_3$ by $b_1$.

Let us assume that $\beta'' \neq 0$. This assumption enables us to rewrite equation~\eqref{zhiber78} in the form
\[
-c^2 \frac{1}{\beta''(u_x)} \left( \frac{\beta''(u_x)}{\beta'^2(u_x)} \right)' = c_4 \gamma'(u_y).
\]
Since $u_x$, $u_y$ are regarded as independent variables, the above equation is equivalent to the system
\begin{gather}
-c^2 \frac{1}{\beta''} \left( \frac{\beta''}{\beta'^2} \right)' = c_5, \qquad c_4 \gamma' = c_5. \label{zhiber80}
\end{gather}
If $c_4 = 0$ then $c_5 = 0$, which yields $c = 0$ or $\beta'' / \beta'^2 = - c_6 \neq 0$. The last equation implies
\[
\beta(u_x) = \frac{1}{c_6} \ln(c_6 u_x + c_7).
\]
Substituting this function into equation~\eqref{zhiber78} and using $c_4 = 0$ we can def\/ine the function $\gamma$ as
\[
\gamma(u_y) = \frac{1}{c_8} \ln(c_8 u_y + c_9),
\]
and the following  
equations result in 
\begin{gather*}
u_{xy} = c(c_6u_x + c_7)(c_8 u_y + c_9), \qquad v_{xy} = 0, \\
v = \frac{1}{c_6} \ln(c_6 u_x + c_7) + \frac{1}{c_8} \ln (c_8 u_y + c_9) + c_3u,
\end{gather*}
where $cc_6 + cc_8 + c_3 =0$, $c_3 \neq 0$.
We use the transformations $v/c_3 \rightarrow v$, $x / c_6 \rightarrow x$, and $y / c_8 \rightarrow y$. Replacing $1 / (c_3c_6)$ by $a_1$, $1/ (c_3c_8)$ by $b_1$, and $cc_6c_8$ by $a$, we get~\eqref{zhiber42}. If $c = 0$ then $c_5 = c_4 = 0$, and we obtain~\eqref{zhiber43}.

Let us turn back to the system \eqref{zhiber80}. Given the assumption $c_4 \neq 0$, this enables us to f\/ind the function $\gamma$,
\[
\gamma(u_y) = \frac{c_5}{c_4} u_y + c_6.
\]
We also have an ordinary dif\/ferential equation def\/ining the function $\beta$,
\[
-c^2 \frac{\beta''}{\beta'^2} = c_5 \beta' + c_7.
\]
Rewriting 
equation~\eqref{zhiber78} by using these equations we get $c_7 + c c_3 = 0$ and, therefore,
\[
u_{xy} = \frac{c c_4}{c_5 \beta'(u_x)}, \qquad v = \frac{c_5}{c_4} u_y + \beta(u_x) + c_3 u, \qquad v_{xy} = c_4,
\]
where $ -c^2 \beta'' / \beta'^2 = c_5 \beta' + c_7$, $ c_7 + c c_3 = 0$, and $c_4 c_5 \neq 0$. Clearly, this case coincides with equation~\eqref{zhiber41} up to the permutation of $x$ and $y$.
\end{proof}

\begin{lemma}\label{lemma2}
Assume that \eqref{zhiber27} is satisfied and $\mu'(u) \neq 0$. Then equations~\eqref{u_xy}, \eqref{v_xy}, and~\eqref{phi} take one of the following forms:
\begin{alignat}{6}
& u_{xy} = \mu(u) u_x u_y, \qquad && v_{xy} = 0, \qquad && v = c_1 \ln u_x + c_2 \ln u_y + \alpha(u), & \label{zhiber81}
\intertext{where the functions $\mu$ and $\alpha$ satisfy
$\mu'(c_1 + c_2) + \mu^2 (c_1 + c_2) + \alpha'' + \alpha' \mu = 0$, $\mu' \neq 0$;}
& u_{xy} = \mu(u)u_x u_y, \qquad && v_{xy} = \exp v, \qquad && v = \ln (u_x u_y) + \alpha(u), & \label{zhiber82}
\end{alignat}
where $\mu$ and $\alpha$ satisfy
$ 2 \mu' + 2 \mu^2 + \alpha'' + \alpha' \mu = \exp \alpha$,
up to the point transformations $u \rightarrow \theta(u)$, $v \rightarrow \kappa(v)$, $x \rightarrow \xi x$, and $y \rightarrow \eta y$, where $\xi$ and $\eta$ are arbitrary constants. Here $c_1$ and $c_2$ are nonzero constants.
\end{lemma}

\begin{proof} Condition \eqref{zhiber27} allows us to determine the functions $\beta$ and $\gamma$ as
\[
\beta(u_x) = c_1 \ln u_x, \qquad \gamma(u_y) = c_2 \ln u_y.
\]
Using these equations \eqref{zhiber23} can be written in the form 
\begin{gather}
 \mu'(u) u_x u_y \left( \frac{1}{c_2} + \frac{1}{c_1} \right) + \frac{\mu^2(u)u_x u_y}{c_1 c_2} \left( \frac{1}{c_2} + \frac{1}{c_1} \right) + \alpha''(u) u_x u_y + \alpha'(u) \mu(u) \frac{u_x u_y}{c_1 c_2} \nonumber\\
 \qquad {} = F \bigl( c_1 \ln u_x + c_2 \ln u_y + \alpha(u) \bigr).\label{zhiber83}
\end{gather}
If we apply the operator $\frac{\partial}{\partial u_x}$ to both sides of equation~\eqref{zhiber83}, we obtain
\[
\mu'(u) u_y \left( \frac{1}{c_2} + \frac{1}{c_1} \right) + \frac{\mu^2(u) u_y}{c_1 c_2} \left( \frac{1}{c_2} + \frac{1}{c_1} \right) + \alpha''(u) u_y + \alpha'(u) \mu(u) \frac{ u_y}{c_1 c_2} = \frac{F' c_1}{u_x}.
\]
Comparing the above equation with equation~\eqref{zhiber83} we notice that $F = c_1 F'$. Similarly, dif\/fe\-ren\-tiating equation~\eqref{zhiber83} with respect to $u_y$ we deduce that $F = c_2F'$. These equations yield $F' = 0$ or $c_2 = c_1$.

If $F' = 0$, equation~\eqref{zhiber83} takes the form
\[
u_x u_y \left( \mu'(u) \left( \frac{1}{c_2} + \frac{1}{c_1} \right) + \frac{\mu^2}{c_1 c_2} \left( \frac{1}{c_2} + \frac{1}{c_1} \right) + \alpha'' + \frac{\alpha' \mu}{c_1 c_2} \right) = c.
\]
 Since $u$, $u_x$, and $u_y$ are regarded as independent variables and the functions $\mu$ and $\alpha$ are functions depending on $u$, we conclude that $c = 0$. Consequently, we obtain the equations
\[
u_{xy} = \frac{\mu(u) u_x u_y}{c_1 c_2}, \qquad v_{xy} = 0, \qquad v = c_1 \ln u_x + c_2 \ln u_y + \alpha(u),
\]
where \[\mu' \left( \frac{1}{c_2} + \frac{1}{c_1} \right) + \frac{\mu^2}{c_1 c_2} \left( \frac{1}{c_2} + \frac{1}{c_1} \right) + \alpha'' + \frac{\alpha' \mu}{c_1 c_2} = 0.\]
Finally, replacing $\mu / c_1 c_2$ by $\mu$ we get equation~\eqref{zhiber81}.

If we replace $c_2$ with $c_1$, we determine $F = c_3 \exp (v/ c_1)$. Equation \eqref{zhiber83} turns into
\[
\frac{2 \mu' u_x u_y}{c_1} + \frac{2 \mu^2 u_x u_y}{c^3_1} + \alpha'' u_x u_y + \frac{\alpha' \mu u_x u_y}{c^2_1} = c_3 u_x u_y \exp( \alpha(u) / c_1).
\]
Thus, the following equations appear
\[
u_{xy} = \frac{1}{c^2_1} \mu(u) u_x u_y, \qquad v_{xy} = c_3 \exp(v / c_1), \qquad v = c_1 \ln u_x u_y + \alpha(u),
\]
where
\[
\frac{2 \mu'}{c_1} + \frac{2 \mu^2}{c^3_1} + \alpha'' + \frac{\alpha' \mu}{c^2_1} = c_3 \exp(\alpha / c_1).
\]
First, we redenote $\mu / c^2_1$ by $\mu$ and $\alpha / c_1$ by $\alpha $. Second, use the transformation $v \rightarrow c_1 v$ and then the shift $v \rightarrow v - \ln c$. Finally, replace $\alpha + \ln c$ by $\alpha$, $c_3 / c$ by $c_1$, and obtain the equations~\eqref{zhiber82}.
\end{proof}

\begin{lemma}\label{lemma3}
Assume that condition \eqref{zhiber30} is satisfied but \eqref{zhiber26} and \eqref{zhiber27} are not. Then equations~\eqref{u_xy}, \eqref{v_xy}, and~\eqref{phi} take one of the following forms:
\begin{alignat}{6}
& u_{xy} = u, \qquad && v_{xy} = v, \qquad && v = c_1 u_y + c_2 u_x + c_3 u; & \label{zhiber84}\\
& u_{xy} = \mu(u) (u_y + b)u_x,\qquad && v_{xy} = \exp v, \qquad && v = \ln(u_y + b) + \ln u_x + \alpha(u), & \label{zhiber85}
\intertext{where the functions $\mu$ and $\alpha$ satisfy $2 \mu' + 2\mu^2 + \alpha'' + \alpha' \mu = \exp \alpha$, $2 \mu^2 + \mu' + \alpha' \mu = \exp \alpha$;}
& u_{xy} = \mu(u)(u_y + b) u_x, \qquad && v_{xy} = 0, \qquad && v = c_2 \ln(u_y + b) + c_1 \ln u_x + \alpha(u),& \label{zhiber86}
\intertext{where $\mu$ and $\alpha$ satisfy $(\mu' + \mu^2)(c_1 + c_2) + \alpha'' + \alpha' \mu = 0$, $c_1 \mu' + \mu^2 (c_1 + c_2) + \alpha' \mu = 0$;}
& u_{xy} = \mu(u)u_x, \qquad && v_{xy} = 0, \qquad && v = u_y - \ln u_x + \alpha(u), & \label{zhiber87}
\end{alignat}
where $\mu$ and $\alpha$ satisfy $\alpha'' + \mu' = 0$, $\mu^2 - \mu' + \alpha' \mu = 0$,
up to the point transformations $u \rightarrow \theta(u)$, $v \rightarrow \kappa(v)$, $x \rightarrow \xi x$ and $y \rightarrow \eta y$, where $\xi$ and $\eta$ are arbitrary constants. Here $c_3$ is an arbitrary constant, $c_1$, $c_2$, and $b$ are nonzero constants.
\end{lemma}

\begin{proof} Condition \eqref{zhiber30} implies the following three possibilities for functions $\beta$ and $\gamma$
\begin{alignat}{3}
& \gamma(u_y) = c_1 u_y + c_2, \qquad && \beta(u_x) = c_3 u_x + c_4, & \label{zhiber88}\\
& \gamma(u_y) = - \frac{1}{c_1} \ln (a_1 u_y + b_1), \qquad && \beta(u_x) = -\frac{1}{c_2} \ln (a_2 u_x + b_2), & \label{zhiber89}\\
& \gamma(u_y) = c_1 u_y + c_2, \qquad && \beta(u_x) = -\frac{1}{c_3} \ln (a u_x + b). & \label{zhiber90}
\end{alignat}
According to \eqref{zhiber88}, equation~\eqref{zhiber23} can be written as
\begin{gather}
\frac{\mu'(u) u_y}{c_3} + \frac{\mu'(u) u_x}{c_1} + \alpha''(u) u_x u_y + \frac{\alpha'(u) \mu(u)}{c_1 c_3} = F(c_1 u_y + c_3 u_x + \alpha(u)).   \label{zhiber91}
\end{gather}
Applying the operators $\frac{\partial}{\partial u_x}$ and $\frac{\partial}{\partial u_y}$ to both sides of \eqref{zhiber91} gives
\[
\frac{\mu'}{c_1} + \alpha'' u_y = F' c_3, \qquad \frac{\mu'}{c_3} + \alpha'' u_x = F' c_1.  \label{zhiber92}
\]
Eliminating $F'$ from the above equations we obtain $\alpha'' (c_1 u_y - c_3 u_x) = 0$. Clearly, we have $\alpha'' = 0$, hence 
$\alpha = c_2 u + c_4$. Furthermore, by using any of the above equations we obtain $F' = \mu' / c_1 c_3$. Consequently,
\[
F(z) = \frac{c_5}{c_1 c_3} z + c_7, \qquad z=c_1u_y+c_3u_x+\alpha(u).
\]
The equation 
\[
\frac{\mu'' u_y}{c_3} + \frac{\mu'' u_x}{c_1} + \frac{c_2 \mu'}{c_1 c_3} = F' c_2
\]
arises after 
the dif\/ferentiation of
 equation~\eqref{zhiber91} with respect to $u$.
Substituting $F' = \mu' / c_1 c_3$ into this equation yields $\mu(u) = c_5 u + c_6$.
Therefore, the equation \eqref{zhiber91} is equivalent to
\[
\frac{c_2 c_6}{c_1 c_3} = \frac{c_5 c_4}{c_1 c_3} + c_7.
\]
Thus, we f\/ind that equations~\eqref{u_xy}, \eqref{v_xy}, and the substitution \eqref{phi} have the forms
\[
u_{xy} = \frac{c_5 u + c_6}{c_1 c_3}, \qquad v_{xy} = \frac{c_5}{c_1 c_3} v + c_7, \qquad
v = c_1 u_y + c_3 u_x + c_2 u + c_4.
\]
 Using the transformations $u + c_6 / c_5 \rightarrow cu$, $v + c_1 c_3c_7 /c_5 \rightarrow cv$ and replacing $c_5 / c_1$ by $c_3$ we get~\eqref{zhiber84}.

Let us discuss the case when the functions $\gamma$ and $\beta$ are of form \eqref{zhiber89}. It turns out that equation~\eqref{zhiber23} takes the form 
\begin{gather}
 -c_2 \mu' u_y \frac{a_2 u_x + b_2}{a_2} - \mu^2 \frac{c_1 c^2_2}{a_1 a_2} (a_2 u_x + b_2) (a_1 u_y + b_1) - \mu^2 \frac{c^2_1 c_2}{a_1 a_2}(a_1 u_y + b_1)(a_2u_x + b_2) \nonumber\\
\qquad{}- c_1 \mu' u_x \frac{a_1 u_y + b_1}{a_1} + \alpha'' u_x u_y + \alpha' \mu \frac{c_1 c_2}{a_1 a_2} (a_2 u_x + b_2)(a_1 u_y + b_1) \nonumber\\
\qquad\quad{}= F \left( -\frac{1}{c_1} \ln(a_1 u_y + b_1) - \frac{1}{c_2} \ln(a_2 u_x + b_2) + \alpha(u) \right).\label{zhiber93}
\end{gather}
Applying the operator $\frac{\partial}{\partial u_x}$ to both sides of equation~\eqref{zhiber93} leads to
\begin{gather*}
-c_2 \mu' u_y - \mu^2 \frac{c_1 c^2_2}{a_1} (a_1 u_y + b_1) - c_1 \mu' \frac{a_1 u_y + b_1}{a_1} - \mu^2 \frac{c^2_1 c_2}{a_1} (a_1 u_y + b_1) \\
\qquad{}+ \alpha'' u_y + \alpha' \mu \frac{c_1 c_2}{a_1} (a_1 u_y + b_1) = F' \left( -\frac{1}{c_2} \right) \frac{a_2}{a_2 u_x + b_2}.
\end{gather*}
The last equation and equation~\eqref{zhiber93} imply
\[
F' \left( -\frac{1}{c_2} \right) - F = -c_1 \mu' \frac{a_1 u_y + b_1}{a_1} \frac{b_2}{a_2} + \alpha'' u_y \frac{b_2}{a_2}.
\]
Similarly, dif\/ferentiating equation~\eqref{zhiber93} with respect to $u_y$ we obtain
\[
F' \left( -\frac{1}{c_1} \right) - F = -c_2 \mu' \frac{a_2 u_x + b_2}{a_2} \frac{b_1}{a_1} + \alpha'' u_x \frac{b_1}{a_1}.
\]
To eliminate $u_x$ and $u_y$ we apply the operators $\frac{\partial}{\partial u_x}$ and $\frac{\partial}{\partial u_y}$ to the two above equations, respectively. We get
\[
F'' \left( - \frac{1}{c_2} - F' \right) = 0, \qquad F'' \left( -\frac{1}{c_1} - F' \right) = 0,
\]
therefore $F''(c_2 - c_1) = 0$.

Assuming that $c_1 = c_2 =c$ we def\/ine $F$ as follows
\[
F(z) = -\frac{1}{c} \exp(-c z + c_7) + c_8.
\]
Substituting the above function $F$ into equation~\eqref{zhiber93} we get
\begin{gather*}
 -\mu' u_y c \left( u_x + \frac{b_2}{a_2} \right) - 2 \mu^2 \frac{c^3}{a_1 a_2} (a_2 u_x + b_2) (a_1 u_y + b_1) - \mu' u_x c \left( u_y + \frac{b_1}{a_1} \right) + \alpha'' u_x u_y \\
\qquad{} + \alpha' \mu \frac{c^2}{a_1 a_2} (a_2 u_x + b_2) (a_1 u_y + b_1) = - \frac{1}{c} (a_2 u_x + b_2) (a_1 u_y + b_1) \exp ( -c \alpha + c_7 ) + c_8.
\end{gather*}
Since $u$, $u_x$, and $u_y$ are considered as independent variables, the above equation is equivalent to the following system
\begin{subequations} \label{zhiber97}
\begin{align}
& \displaystyle - 2c \mu' - 2 \mu^2 c^3 + \alpha'' + \alpha' \mu c^2 = -\frac{a_1 a_2}{c} \exp (-c \alpha + c_7), \\
& \displaystyle -2\mu^2 \frac{c^3}{a_1}b_1 - \mu' c \frac{b_1}{a_1} + \alpha' \mu c^2 \frac{b_1}{a_1} = - \frac{1}{c} a_2 b_1 \exp (-c \alpha + c_7), \\
& \displaystyle -\mu' c \frac{b_2}{a_2} - 2 \mu^2 c^3 \frac{b_2}{a_2} + \alpha' \mu c^2 \frac{b_2}{a_2} = - \frac{1}{c} a_1 b_2 \exp(-c \alpha + c_7),
 \\
& \displaystyle -2 \mu^2 c^3 \frac{b_1 b_2}{a_1 a_2} + \alpha' \mu c^2 \frac{b_1 b_2}{a_1 a_2} = - \frac{1}{c} b_1 b_2 \exp(-c \alpha + c_7) + c_8.
\end{align}
\end{subequations}
Note that $(b_1, b_2) \neq (0, 0)$. Otherwise, condition \eqref{zhiber27} is true, which contradicts the assumption of the lemma. 
If $b_2 = 0$, $b_1 \neq 0$ then $c_8 = 0$ and
\begin{gather}\label{zhiber97.5}
u_{xy} = \frac{\mu(u)c^2}{a_1} (a_1 u_y + b_1)u_x, \qquad v_{xy} = -\frac{1}{c} \exp(-cv+c_7),\nonumber \\
v = -\frac{1}{c} \ln (a_1 u_y + b_1) - \frac{1}{c} \ln (a_2 u_x) + \alpha(u),\end{gather}
where the functions $\mu$ and $\alpha$ satisfy the following equations
\begin{gather*}
-2c\mu' - 2\mu^2 c^3 + \alpha'' + \alpha' \mu c^2 = -\frac{a_1 a_2}{c} \exp(-c \alpha + c_7), \\
-2 \mu^2 c^3 - \mu' c + \alpha' \mu c^2 = - \frac{a_1 a_2}{c} \exp(-c \alpha + c_7).
\end{gather*}
Applying the transformation $ -cv + c_7\rightarrow v $ and redenoting $-c \alpha + c_7 + \ln (a_1 a_2)$ by $\alpha$, $\mu c^2$ by $\mu$ and $b_1 / a_1$ by $b$,
we transform \eqref{zhiber97.5} into \eqref{zhiber85}. It is not hard to prove that system \eqref{zhiber97} has no solutions if $b_1b_2 \neq 0$.

Let us suppose that $F'' = 0$, hence $F(z) = cz + p$, where $c$ and $p$ are arbitrary constants. In this case
 equation~\eqref{zhiber93} is represented as 
\begin{gather*}
  -c_2 \mu' u_y \left( u_x + \frac{b_2}{a_2} \right) - \mu^2 c_1 c^2_2 \left( u_x + \frac{b_2}{a_2} \right) \left( u_y + \frac{b_1}{a_1} \right) - c_1 \mu' u_x \left( u_y + \frac{b_1}{a_1} \right) + \alpha'' u_x u_y\\
\qquad{} - \mu^2 c^2_1 c_2 \left( u_x + \frac{b_2}{a_2} \right) \left( u_y + \frac{b_1}{a_1} \right) + \alpha' \mu c_1 c_2 \left( u_x + \frac{b_2}{a_2} \right) \left( u_y + \frac{b_1}{a_1} \right) \\
\qquad\quad{} = c \left( - \frac{1}{c_1} \ln (a_1 u_y + b_1) - \frac{1}{c_2} \ln(a_2 u_x + b_2) + \alpha(u) \right) + p.
\end{gather*}
It is clear that the coef\/f\/icients at $\ln (a_1 u_y + b_1)$ and $\ln(a_2 u_x + b_2)$ are equal to zero, i.e.\ $c = 0$. Since $u$, $u_x$, and $u_y$ are regarded as independent variables, the above equation is equivalent to the system
\begin{gather*}
 -c_2 \mu' - \mu^2 c_1 c^2_2 - c_1 \mu' - \mu^2 c^2_1 c_2 + \alpha'' + \alpha' \mu c_1 c_2 = 0, \\
 -\mu^2 c_1 c^2_2 \frac{b_1}{a_1} - c_1 \mu' \frac{b_1}{a_1} - \mu^2 c^2_1 c_2 \frac{b_1}{a_1} + \alpha' \mu c_1 c_2 \frac{b_1}{a_1} = 0,\\
 -c_2 \mu' \frac{b_2}{a_2} - \mu^2 c_1 c^2_2 \frac{b_2}{a_1} - \mu^2 c^2_1 c_2 \frac{b_2}{a_1} + \alpha' \mu c_1 c_2 \frac{b_2}{a_2} = 0, \\
 {-\mu^2 c_1 c^2_2 - \mu^2 c^2_1 c_2 + \alpha' \mu c_1 c_2} \frac{b_1 b_2}{a_1 a_2} = p.
\end{gather*}
Note that $(b_1,b_2) \neq (0, 0)$. Otherwise, condition \eqref{zhiber27} is satisf\/ied, which contradicts the assumption of the lemma.
If $b_2 = 0$, $b_1 \neq 0$ then $p = 0$ and
\begin{gather*}
u_{xy} = \mu(u) \frac{c_1 c_2}{a_1} (a_1 u_y + b_1)u_x, \qquad v_{xy} = 0, \\
v = -\frac{1}{c_1}\ln(a_1 u_y + b_1) - \frac{1}{c_2} \ln(a_2 u_x) + \alpha(u),
\end{gather*}
where the functions $\mu$ and $\alpha$ satisfy the equations
\begin{gather*}
 \mu'(c_1 + c_2) + \mu^2 c_1 c_2 (c_1 + c_2) - \alpha'' - \alpha' \mu c_1 c_2 = 0,
 \\
  c_1 \mu' + \mu^2 c_1 c_2 (c_1 + c_2) - \alpha' \mu c_1 c_2 = 0.
\end{gather*}
We replace $c_1 c_2 \mu$ by $\mu$, $-c_1 c_2 \alpha + c_2 \ln a_1 + c_1 \ln a_2$ by $\alpha$.
Using the transformation $ v \rightarrow -v/ (c_1 c_2)$ and redenoting $b_1 / a_1$ by $b$ we transform the above equations into~\eqref{zhiber86}. If $b_1 b_2 \neq 0$ then the last system has no solutions.

Let us suppose that the functions $\gamma$ and $\beta$ are given by \eqref{zhiber90}. We rewrite equation~\eqref{zhiber23} using \eqref{zhiber90},
\begin{gather}
 -\frac{c_2}{a} \mu' u_y (au_x + b) + \frac{c^2_2}{a c_1} (a u_x + b) + \frac{1}{c_1} \mu' u_x + \alpha'' u_x u_y + \alpha' \mu (a u_x + b) \left( - \frac{c_2}{c_1 a} \right) \nonumber\\
\qquad{} = F \left( c_1 u_y - \frac{1}{c_2} \ln (au_x + b) + \alpha(u) \right).\label{zhiber98}
\end{gather}
Applying the operators $\frac{\partial}{\partial u_x}$ and $\frac{\partial}{\partial u_y}$ to both sides of equation~\eqref{zhiber98} we obtain
\begin{gather}\label{zhiber99}
-c_2 \mu' u_y + \frac{c^2_2}{c_1} \mu^2 + \frac{1}{c_1} \mu' + \alpha'' u_y - \frac{c_2}{c_1} \alpha' \mu = F' \left( - \frac{1}{c_2} \right) \frac{a}{a u_x + b}, \\
-\frac{c_2}{a} \mu' (au_x + b) + \alpha'' u_x = F' c_1. \label{zhiber100}
\end{gather}

If $F' = 0$ then we obviously get $F = c_3$ and
\[
-c_2 \mu' + \alpha'' = 0, \qquad c^2_2 \mu^2 + \mu' - c_2 \alpha' \mu = 0, \qquad \mu' b = 0.
\]
We analyze equation~\eqref{zhiber98} based on these equations and f\/ind that $c_3 = 0$. It allows us to determine equations~\eqref{u_xy}, \eqref{v_xy}, and \eqref{phi} as follows
\[
u_{xy} = -\frac{c_2}{c_1} \mu(u) u_x, \qquad v_{xy} = 0 , \qquad v = c_1 u_y - \frac{1}{c_2} \ln (au_x) + \alpha(u),
\]
where the functions $\mu$ and $\alpha$ satisfy
\[
\alpha'' = c_2 \mu', \qquad c^2_2 \mu^2 + \mu' - c_2 \alpha' \mu =0.
\]
Point transformations enable us to represent the above equations in form \eqref{zhiber87}.

Assuming that $F' \neq 0$ we can eliminate $F'$ from equations~\eqref{zhiber99} and \eqref{zhiber100}
\begin{gather*}
 c^2_2 \left( \frac{a u_x + b}{a} \right) \mu' u_y - \frac{c^3_2}{c_1} \left( \frac{a u_x + b}{a} \right) \mu^2 - \frac{c_2}{c_1} \left( \frac{a u_x + b}{a} \right) \mu' - c_2 \left( \frac{au_x + b}{a} \right) \alpha'' u_y \\
\qquad{}  + \frac{c^2_2}{c_1} \alpha' \mu \left( \frac{a u_x + b}{a} \right) = -\frac{c_2}{c_1 a} (a u_x + b) \mu' + \frac{\alpha''}{c_1}u_x.
\end{gather*}
Recall that variables $u$, $u_x$, and $u_y$ are considered as independent. Hence, the above equation is equivalent to the system
\begin{subequations} \label{zhiber101}
\begin{gather}
 c^2_2 \mu' - c_2 \alpha'' = 0,\\
 -\frac{c^3_2}{c_1} \mu^2 - \frac{c_2}{c_1} \mu' + \frac{c^2_2}{c_1} \alpha' \mu = - \frac{c_2}{c_1} \mu' + \frac{\alpha''}{c_1}, \\
 \frac{c^2_2 b}{a} \mu' - c_2 \frac{b}{a} \alpha'' =0,\\
 - \frac{c^3_2}{c_1} \frac{b}{a} \mu^2 + \frac{c^2_2}{c_1} \alpha' \mu \frac{b}{a} = 0.
\end{gather}
\end{subequations}
If $b = 0$, we transform equation~\eqref{zhiber98} into
\[
-c_2 \mu' u_x u_y + \frac{c^2_2}{c_1} \mu^2 u_x + \frac{1}{c_1} \mu' u_x + \alpha'' u_x u_y - \frac{c_2}{c_1} \alpha' \mu u_x = F \left( c_1 u_y - \frac{1}{c_2} \ln (a u_x) + \alpha(u) \right).
\]
Dif\/ferentiating this equation with respect to $u_x$ we obtain
\[
\frac{c^2_2}{c_1} \mu^2 - c_2 \mu' u_y + \frac{1}{c_1} \mu' + \alpha'' u_y - \frac{c_2}{c_1} \alpha' \mu = -\frac{1}{c_2} F' \frac{1}{u_x}.
\]
One can notice that these two equations imply $F + F'/c_2 = 0$ or $F(z) = c_3 \exp (-c_2 z)$. Consequently, we get
\[
-c_2 \mu' u_x u_y + \frac{c^2_2}{c_1} \mu^2 u_x + \frac{1}{c_1} \mu' u_x + \alpha'' u_x u_y - \frac{c_2}{c_1} \alpha' \mu u_x = c_3 \exp(-c_2 c_1 u_y) a u_x \exp(\alpha).
\]
This equation is not realized because of the given assumptions $c_3 \neq 0$ and $a \neq 0$.

Now, it remains only to consider the case when $b \neq 0$. System \eqref{zhiber101} takes the form
\[
c_2 \mu' - \alpha'' = 0, \qquad -c^3_2 \mu^2 + c^2_2 \alpha' \mu = \alpha'', \qquad -c_2 \mu^2 + \alpha' \mu = 0.
\]
These equations imply that $\mu' = 0$, which contradicts the given assumptions of the lemma. 
\end{proof}

\begin{lemma}\label{lemma4}
Suppose that condition \eqref{zhiber28} holds but \eqref{zhiber26}, \eqref{zhiber27}, and \eqref{zhiber30} do not. Then equations~\eqref{u_xy}, \eqref{v_xy}, and \eqref{phi} take one of the following forms:
\begin{alignat}{6}
&  u_{xy} = \frac{\mu(u) u_x}{\gamma'(u_y)}, \qquad && v_{xy} = 0, \qquad && v = \ln u_x + \gamma(u_y) + \alpha(u), & \label{zhiber102}
\intertext{where $c_3 + \frac{\gamma''}{\gamma'^2} + c_4 \gamma' u_y = 0$, $\alpha'' + \mu' + c_4 \mu^2 = 0$, and $c_3 \mu^2 + \mu' + \mu^2 + \alpha' \mu =0$;}
& u_{xy} = \frac{u_x}{(au + b) \gamma'(u_y)}, \qquad && v_{xy} = \exp v,\qquad &&&\nonumber\\
 &&& v = \ln u_x + \gamma(u_y) - 2 \ln(au + b) + \ln(-c_5),\hspace*{-200mm} &&&  \label{zhiber103}
\end{alignat}
where $c_3 + \frac{\gamma''}{\gamma'^2} + c_4 \gamma' u_y = c_5 \gamma' \exp \gamma$, $c_3 + 1 -3a = 0,$ and $c_4 + 2a^2 - a = 0$,
up to the point transformations $u \rightarrow \theta(u)$, $v \rightarrow \kappa(v)$, $x \rightarrow \xi x$, and $y \rightarrow \eta y$, where $\xi$ and $\eta$ are arbitrary constants. Here $c_3$, $c_4$ are arbitrary constants, $c_5 \neq 0$, and $(a, b) \neq (0, 0)$.
\end{lemma}

\begin{proof}
According to \eqref{zhiber28}, the function $\beta$ is of the form $\beta = c_1 \ln u_x + c_2$. Without loss of generality, we may set $\beta = c_1 \ln u_x$. Substituting $\beta$ into equation~\eqref{zhiber23} we obtain 
\begin{gather}
\frac{\alpha'\mu u_x}{c_1 \gamma'} - \frac{\mu^2 u_x}{c_1 \gamma'} \left( \frac{\gamma''}{\gamma'^2} - \frac{1}{c_1} \right) + \alpha'' u_x u_y + \mu' \left( \frac{u_x}{\gamma'} + \frac{u_x u_y}{c_1} \right) = F(\alpha + \beta + \gamma). \label{zhiber105}
\end{gather}
Applying the operator $\frac{\partial}{\partial u_x}$ to both sides of \eqref{zhiber105} leads to
\begin{gather}
\frac{\alpha' \mu}{c_1 \gamma'} - \frac{\mu^2}{c_1 \gamma'} \left( \frac{\gamma''}{\gamma'^2} - \frac{1}{c_1} \right) + \alpha'' u_y + \mu' \left( \frac{1}{\gamma'} + \frac{u_y}{c_1} \right) = F' \left( \frac{c_1}{u_x} \right). \label{zhiber106}
\end{gather}
From equations~\eqref{zhiber105} and \eqref{zhiber106} it follows that $F = F' c_1/u_x$, hence $F(z) = c_2 \exp(z / c_1)$.
By substituting $F$ into equation~\eqref{zhiber105} we get
\[
u_x \left( \frac{\alpha' \mu}{c_1 \gamma'} - \frac{\mu^2}{c_1 \gamma'}\left( \frac{\gamma''}{\gamma'^2} - \frac{1}{c_1}\right) + \alpha'' u_y + \mu' \left( \frac{1}{\gamma'} + \frac{u_y}{c_1} \right) \right) = c_2 u_x \exp(\gamma / c_1) \exp(\alpha / c_1).
\]
This equation can be written in the form
\[
\mu' c_1 + \alpha' \mu + \frac{\mu^2}{c_1} - \mu^2 \frac{\gamma''}{\gamma'^2} + (\alpha'' c_1 + \mu') \gamma' u_y = c_2 c_1 \gamma' \exp(\gamma / c_1) \exp(\alpha / c_1).
\]
Having the f\/ixed value of $u$ we can determine $\gamma$ as a solution of the ordinary dif\/ferential equation
\[
c_3 + \frac{\gamma''}{\gamma'^2} + c_4 \gamma' u_y = c_1 c_5 \gamma' \exp(\gamma / c_1).
\]
Moreover, based on this equation we get 
\begin{gather*}
\alpha' \mu + \frac{\mu^2}{c_1} + c_1 \mu' + c_3 \mu^2 + \gamma u_y \big(c_1 \alpha'' + \mu' + c_4 \mu^2\big)\\
 \qquad{}- c_1 \gamma' \exp(\gamma / c_1) (c_5 \mu^2 + c_2 \exp(\alpha / c_1)) = 0.
\end{gather*}
Note that if $u_y = \kappa \exp(\gamma / c_1)$ then $\gamma = c_1 \ln (u_y / \kappa)$ and $(\gamma' u_y)' = 0$. Since the last equation contradicts the assumption of the lemma, 
we obtain that $u_y$ and $\exp(\gamma / c_1)$ are linearly independent and that is why 
\begin{gather*}
c_1 \alpha'' + \mu' + c_4 \mu^2 = 0, \qquad c_5 \mu^2 + c_2 \exp(\alpha / c_1) = 0, \qquad c_3 \mu^2 + c_1 \mu' + \frac{\mu^2}{c_1} + \alpha' \mu = 0.
\end{gather*}

In order to f\/ind equations~\eqref{u_xy}, \eqref{v_xy}, and \eqref{phi} we f\/irst set $c_5 = 0$, hence $c_2 = 0$ and
\[
u_{xy} = \frac{\mu(u)}{\beta'(u_x)\gamma'(u_y)}, \qquad v_{xy} = 0, \qquad v = \beta(u_x) + \gamma(u_y) + \alpha(u),
\]
where the functions $\beta$ and $\gamma$ are 
solutions of the ordinary dif\/ferential equations
\[
\beta' = \frac{c_1}{u_x}, \qquad c_3 + \frac{\gamma''}{\gamma'^2} + c_4 \gamma' u_y = 0,
\]
and the functions $\mu$ and $\alpha$ satisfy the equations
\[
c_1 \alpha'' + \mu' + c_4 \mu^2 = 0, \qquad c_3 \mu^2 + c_1 \mu' + \frac{\mu^2}{c_1} + \alpha' \mu = 0.
\]
We use the transformation $v \rightarrow c_1 v$. Next, we redenote $\alpha / c_1$ by $\alpha$, $\gamma / c_1$ by $\gamma$, and $\mu / c^2_1 $ by $ \mu$. Finally, after replacing $c_4 c^2_1$ by $c_4$ and $c_1 c_3$ by $c_3$, \eqref{zhiber102} is obtained.

If $c_5 \neq 0$ then we get
\[
u_{xy} = \frac{\mu(u)}{\beta'(u_x) \gamma'(u_y)}, \qquad v_{xy} = c_2 \exp(v / c_1), \qquad v = \beta(u_x) + \gamma(u_y) + \alpha(u),
\]
where the functions $\beta$ and $\gamma$ are the solutions of the ordinary dif\/ferential equations
\[
\beta' = \frac{c_1}{u_x}, \qquad c_3 + \frac{\gamma''}{\gamma'^2} + c_4 \gamma' u_y = c_1 c_5 \gamma' \exp(\gamma / c_1),
\]
and the functions $\alpha$ and $\mu$ are given by the equations 
\begin{gather*}
\alpha = 2 c_1 \ln(-2c_1) - 2c_1 \ln \left( -\frac{2}{3} \sqrt{- \frac{c_2}{c_5}} \frac{c_3 c_1 + 1}{c_1} u + c_6 \right), \\
\mu = \sqrt{-\frac{c_2}{c_5}} \left( \frac{-2c_1}{-\frac{2}{3} \sqrt{-\frac{c_2}{c_5}} \left( \frac{c_3 c_1 + 1}{c_1} \right)u + c_6} \right),\\
\frac{2}{9} \left( \frac{c_3c_1 +1}{c_1} \right)^2 - \frac{1}{3} \left( \frac{c_3 c_1 + 1}{c^2_1} \right) + c_4 = 0.
\end{gather*}
After point transformations we get \eqref{zhiber103}.
\end{proof}

\begin{lemma}\label{lemma5}
Suppose that condition \eqref{zhiber31} holds but \eqref{zhiber26}--\eqref{zhiber30} do not. Then equations~\eqref{u_xy}, \eqref{v_xy}, and \eqref{phi} take one of the following forms:
\begin{alignat}{6}
& u_{xy} = - \frac{1}{u \beta'(u_x) \gamma'(u_y)}, \qquad && v_{xy} = 0, \qquad && v = \beta(u_x) + \gamma(u_y), & \label{zhiber108}
\intertext{where $\frac{\beta''}{\beta'^2} = u_x \beta' + c_1$, $\frac{\gamma''}{\gamma'^2} = u_y \gamma' - c_1$;}
& u_{xy} = \frac{\mu(u)}{\beta'(u_x) \gamma'(u_y)}, \qquad && v_{xy} = \exp v, \qquad && v = \beta(u_x) + \gamma(u_y) + \alpha(u),\hspace*{-50mm}& \label{zhiber109}
\intertext{where $ u_x + \frac{1}{\beta'(u_x)} = \exp (\beta)$, $u_y + \frac{1}{\gamma'(u_y)} = \exp \gamma$, $\alpha'' =\exp \alpha$, and $\mu = (\exp \alpha) / \alpha'$;}
& u_{xy} = \frac{\mu(u)}{\beta'(u_x) \gamma'(u_y)}, \qquad && v_{xy} = \exp v, \qquad && v = \beta(u_x) + \gamma(u_y) + \alpha(u),\hspace*{-50mm}& \label{zhiber110}
\intertext{where
$-c u_x + \frac{1}{\beta'(u_x)} = \exp \beta$, $-c u_y + \frac{1}{\gamma'(u_y)} = \exp \gamma$,
$\alpha' \mu + 2 \mu^2 (c+1) = \exp \alpha$, $\alpha'^2 = 2 c^2 \exp \alpha$, $c = - \frac{1}{2}, -2$;}
& u_{xy} = \frac{\mu(u)}{\beta'(u_x)\gamma'(u_y)}, \qquad && v_{xy} = \exp v + \exp (-v), \qquad && v = \beta(u_x) + \gamma(u_y) + \alpha(u),\hspace*{-50mm}& \label{zhiber111}
\intertext{where
$A_1 \exp \beta + B_1 \exp (-\beta) = u_x$, $A_2 \exp \gamma + B_2 \exp(-\gamma) = u_y$,
$\alpha'' = \frac{1}{4} \left( \frac{\exp(-\alpha)}{B_1 B_2} + \frac{\exp \alpha}{A_1 A_2} \right)$,  $\mu = \frac{\alpha''}{\alpha'}$;}
& u_{xy} = \frac{\mu(u)}{\beta'(u_x) \gamma'(u_y)}, \qquad && v_{xy} = \exp v + \exp(-2 v), \qquad && v = \beta(u_x) + \gamma(u_y) + \alpha(u),\hspace*{-50mm}& \label{zhiber112}
\end{alignat}
where
$A_1 \exp \beta + B_1 \exp (-2 \beta) = u_x$, $A_2 \exp \gamma + B_2 \exp (-2 \gamma) = u_y$,
$\alpha'^2 = \frac{2}{9} \left( \frac{4 \exp \alpha}{A_1 A_2} - \frac{1}{2} \frac{\exp(-2 \alpha)}{B_1 B_2} \right)$,
$-2\mu^2 + \alpha' \mu - \frac{1}{9} \left( \frac{\exp \alpha}{ A_1 A_2} + \frac{\exp(-2\alpha)}{B_1 B_2} \right) = 0$,
up to the point transformations $u \rightarrow \theta(u)$, $v \rightarrow \kappa(v)$, $x \rightarrow \xi x$, and $y \rightarrow \eta y$, where $\xi$ and $\eta$ are arbitrary constants. Here $A_1$, $A_2$, $B_1$, and $B_2$ are nonzero constants.
\end{lemma}

\begin{proof}
Considering that $u_x$ and $u_y$ are independent variables, equation~\eqref{zhiber31} yields
\[
\frac{ (u_x\beta')' }{\left( \dfrac{\beta''}{\beta'^2} \right)'} = c, \qquad \frac{ (u_y\gamma')' }{ \left( \dfrac{\gamma''}{\gamma'^2} \right)' } = c, \qquad c \neq 0.
\]
Integrating these equations we obtain
\begin{gather}
\frac{\beta''}{\beta'^2} = c u_x \beta' + c_1, \qquad \frac{\gamma''}{\gamma'^2} = c u_y \gamma' + c_2. \label{zhiber113}
\end{gather}
According to \eqref{zhiber113}, equation~\eqref{zhiber23} is rewritten in the form 
\begin{gather} \label{zhiber114}
\frac{1}{\beta'} \left(\! \frac{\alpha' \mu}{ \gamma'} - \frac{\mu^2}{\gamma'} \big(c_1 + c u_y \gamma' + c_2\big) + \mu' u_y \right)
 + u_x \!\left(\! -\frac{c \mu^2}{\gamma'} + \alpha'' u_y + \frac{\mu'}{\gamma'} \right) = F(\alpha + \beta + \gamma).\!\!\!\!
\end{gather}
Having f\/ixed values of $u$ and $u_y$ we can def\/ine that $F(\beta + c_3) = c_4 u_x + c_5 / \beta'$. Without loss of generality, we redenote $\beta + c_3$ by $\beta$, therefore
\begin{gather}
F(\beta) = c_4 u_x + \frac{c_5}{\beta'}. \label{zhiber115}
\end{gather}
Applying the operator $\frac{\partial}{\partial u_x}$ to both sides of equation~\eqref{zhiber115} and using \eqref{zhiber113} we obtain
\[
F'(\beta) = -cc_5 u_x + \frac{c_4 - c_1 c_5}{\beta'}.
\]
We dif\/ferentiate this equation with respect to $u_x$,
\[
F''(\beta) = -c(c_4 - c_1 c_5) u_x - \frac{cc_5 +c_1 (c_4 - c_1 c_5)}{\beta'}.
\]
The above three equations allow us to establish that the function $F$ satisf\/ies the ordinary dif\/ferential equation
\begin{gather}
F'' = c_7 F' + c_8 F. \label{zhiber116}
\end{gather}
Equation \eqref{zhiber116} possesses two families of solutions
\[
F(v) = A_1 \exp(\sigma_1 v) + B_1 \exp(\sigma_2 v), \qquad \sigma_1 \neq \sigma_2,
\]
and 
\[
F(v) = (A_2 + B_2 v) \exp(\sigma v).
\]
Setting def\/inite values of the constants $A_i$, $B_i$, where $i=1,2$, we obtain that the function $F$ can take only one of the following forms
\begin{gather}
F(v) = 0,    \label{zhiber117}\\
F(v) = 1,    \label{zhiber118}\\
F(v) = v,    \label{zhiber119}\\
F(v) = v \exp v,   \label{zhiber120}\\
F(v) = \exp v,   \label{zhiber121}\\
F(v) = \exp v + 1,   \label{zhiber122}\\
F(v) = \exp v + \exp (\sigma v). \label{zhiber123}
\end{gather}
From equation~\eqref{zhiber114} by setting dif\/ferent values of $u$ and $u_y$ we obtain a set of equations
\begin{gather}
\alpha_i u_x + \frac{\beta_i}{\beta'(u_x)} = F\left( \beta(u_x) + \gamma_i \right). \label{zhiber124}
\end{gather}
Here $\alpha_i$, $\beta_i$, and $\gamma_i$ are constants, $i= 1, 2, \dots, n$. Thus, we will focus on \eqref{zhiber124}.

Let us assume that $(\alpha_i, \beta_i)$ are linearly dependent vectors. This means that a set of numbers~$\mu_i$ satisfying
\[
(\alpha_i, \beta_i) = \mu_i (\alpha_1, \beta_1), \qquad \mu_1 = 1,
\]
exists.
Using this equation we rewrite \eqref{zhiber124} as
\begin{gather}
\mu_i \left( \alpha_1 u_x + \frac{\beta_1}{\beta'(u_x)} \right) = F(\beta + \gamma_i). \label{zhiber125}
\end{gather}
Now, we will deal with equations~\eqref{zhiber117}--\eqref{zhiber123}.

We begin with \eqref{zhiber117}. In this case we have
\begin{gather}
\mu_i \left( \alpha_1 u_x + \frac{\beta_1}{\beta'(u_x)} \right) = 0 \label{zhiber126}
\end{gather}
from the equation \eqref{zhiber125}. Suppose that $\alpha_1 = \beta_1 = 0$. In equation~\eqref{zhiber114}, we f\/ind
\begin{gather}
\mu' - c \mu^2 + \alpha'' u_y \gamma' = 0, \qquad \alpha' \mu - \mu^2(c_1 + c_2 + c u_y \gamma') + \mu' u_y \gamma' = 0. \label{zhiber127}
\end{gather}

If $\alpha'' = 0$ then $\alpha = \epsilon u + \delta$, hence from \eqref{zhiber127} we have
\[
\mu(u) = - \frac{1}{cu + \kappa}, \qquad
\frac{\epsilon}{cu + \kappa} + \frac{c_1 + c_2}{(cu + \kappa)^2} = 0.
\]
Clearly, the last equation requires $\epsilon =0$ and $c_2 = - c_1$. Thus, we determine equations~\eqref{u_xy}, \eqref{v_xy}, and \eqref{phi} as follows
\[
u_{xy} = \frac{\mu(u)}{\beta'(u_x) \gamma'(u_y)}, \qquad v_{xy} = 0,\qquad v = \beta(u_x) + \gamma(u_y) + \alpha(u),
\]
where
\begin{gather*}
\mu(u) = - \frac{1}{c u + \kappa}, \qquad \alpha(u) = \delta, \qquad \frac{\beta''}{\beta'^2} = c u_x \beta' + c_1, \qquad \frac{\gamma''}{\gamma'^2} = c u_y \gamma' - c_1.
\end{gather*}
We replace $\beta$ by $a \beta$, $ \gamma$ by $a \gamma$. Take the constant $a$ so that $a^2 c \rightarrow 1$. Using the transformations $u + \kappa / c \rightarrow u$, $v - \delta \rightarrow av$ and redenoting $a c_1 \rightarrow c_1$ obtain equation~\eqref{zhiber108}.

Now, assume that $\alpha'' \neq 0$. The equation
\[
u_y \gamma'(u_y) = \frac{c \mu^2 - \mu'}{\alpha''}
\]
arises from \eqref{zhiber127}. Since $u$ and $u_y$ are regarded as independent variables, the last equation leads to $u_y \gamma'(u_y) = \kappa$, where $\kappa$ is a constant. This contradicts the assumption of the lemma. 

Consider the case where $\alpha_1 \beta_1 \neq 0$. We have the equation $\beta'(u_x) = - \beta_1 / (\alpha_1 u_x)$ which results from \eqref{zhiber126}, and it contradicts the assumptions of the lemma. 

Let us discuss the case where $F$ is determined by \eqref{zhiber118}. Rewriting 
\eqref{zhiber125} we have
\[
\mu_i \left( \alpha_1 u_x + \frac{\beta_1}{\beta'(u_x)} \right) = 1.
\]
This equation must be true for every $i=1, 2, \dots$. This requirement implies that
 $\mu_i = 1$, $\alpha_i = \alpha_1$, and $\beta_i = \beta_1$ for every $i$. Taking this into account we def\/ine $\beta'$ as follows:
\begin{gather}
\beta'(u_x) = \frac{\beta_1}{1 - \alpha_1 u_x}.  \label{zhiber128}
\end{gather}
Rewriting 
\eqref{zhiber113} by using \eqref{zhiber128} we see that this case is not realized.

Now, we assume that $F$ is described by \eqref{zhiber119}. Equations~\eqref{zhiber124}, \eqref{zhiber125} are presented in the forms
\[
\alpha_1 u_x + \frac{\beta_1}{\beta'(u_x)} = \beta(u_x) + \gamma_1, \qquad \mu_i \left( \alpha_1 u_x + \frac{\beta_1}{\beta'(u_x)} \right) = \beta(u_x) + \gamma_i.
\]
Consequently,
\[
\beta(u_x) (\mu_i - 1) + \gamma_1 \mu_i - \gamma_i = 0.
\]
It is clear that $\mu_i = 1$, $\gamma_i = \gamma_1$. Hence, $\alpha_i = \alpha_1$, $\beta_i = \beta_1$ for every $i$. So we have
\[
\beta' = \frac{\beta_1}{\beta(u_x) - \alpha_1 u_x + \gamma_1}.
\]
Trying to simplify \eqref{zhiber114} by using this equation gives a contradiction to the assumption of the lemma. 

Concentrate on the case when $F$ satisf\/ies \eqref{zhiber120}. We can rewrite equations~\eqref{zhiber124}, \eqref{zhiber125} as
\[
\alpha_1 u_x + \frac{\beta_1}{\beta'(u_x)} = (\beta + \gamma_1) \exp (\beta + \gamma_1), \qquad \mu_i \left( \alpha_1 u_x + \frac{\beta_1}{\beta'(u_x)} \right) = (\beta + \gamma_i) \exp(\beta + \gamma_i).
\]
Comparing these equations we conclude that
\[
 \left( \beta (\exp \gamma_i - \mu_i \exp \gamma_1) + \gamma_i \exp \gamma_i - \mu_i \gamma_1 \exp \gamma_1 \right)\exp \beta = 0.
\]
Recall that $\beta$ depends on the variable $u_x$, while the remaining terms of the above equations are constants. Hence, we have
\[
\exp \gamma_i - \mu_i \exp \gamma_1 = 0, \qquad \gamma_i \exp \gamma_i - \mu_i \gamma_1 \exp \gamma_1 = 0.
\]
From these equations we obtain $\gamma_i \exp \gamma_i - \gamma_1 \exp \gamma_i = 0$, hence 
$\gamma_i = \gamma_1$ for all $i$. By \eqref{zhiber124} we determine that $\alpha(u) + \gamma(u_y) = \gamma_1$, where $\gamma_1$ is an arbitrary constant. This equation contradicts $\gamma_{u_y} \neq 0$.

Let the function $F$ be def\/ined by \eqref{zhiber121}. From \eqref{zhiber124} we obtain 
\begin{gather}
\alpha_1 u_x + \frac{\beta_1}{\beta'(u_x)} = \exp(\beta + \gamma_1). \label{zhiber129}
\end{gather}
Note that $\beta_1 \neq 0$, otherwise $(\beta' u_x)' = 0$. Redenoting $\beta + \gamma_1$ by $\beta$ we rewrite equation~\eqref{zhiber129} in the form 
\begin{gather}
\alpha_1 u_x + \frac{\beta_1}{\beta'(u_x)} = \exp \beta.  \label{zhiber130}
\end{gather}
From equations~\eqref{zhiber113} and \eqref{zhiber130} we f\/ind that $c = -\alpha_1 / \beta_1$, $c_1 = -1-c$. Now, we rewrite equation~\eqref{zhiber114} based on equation \eqref{zhiber130}
\begin{gather*}
 \frac{1}{\beta'} \exp \beta \left( \frac{\alpha' \mu}{\gamma'} - \frac{\mu^2}{\gamma'} \big(c_1 + c u_y \gamma' + c_2\big) + \mu' u_y \right) + u_x \left( -\frac{c \mu^2}{\gamma'} + \alpha'' u_y + \frac{\mu'}{\gamma'} \right) \\
\qquad{}  - \frac{\alpha_1}{\beta_1} u_x \left( \frac{\alpha' \mu}{\gamma'} - \frac{\mu^2}{\gamma'}\big(c_1 + c u_y \gamma' + c_2\big) + \mu' u_y \right)
= \exp(\alpha + \gamma) \exp \beta.
\end{gather*}
Since $(\beta' u_x)' \neq 0$, $\exp \beta$ and $u_x$ are linearly independent, the above equation is equivalent to the system
\begin{gather*}
\frac{\alpha' \mu}{\gamma'} - \frac{\mu^2}{\gamma'} \big(c_1 + c u_y \gamma' + c_2\big) + \mu' u_y = \exp(\alpha + \gamma) \beta_1,\\
-\frac{\alpha_1}{\beta_1} \left( \frac{\alpha' \mu}{\gamma'} - \frac{\mu^2}{\gamma'}\big(c_1 + c u_y \gamma' + c_2\big) + \mu' u_y \right) + \left( -\frac{c \mu^2}{\gamma'} + \alpha'' u_y + \frac{\mu'}{\gamma'} \right) = 0.
\end{gather*}
Hence, we get
\begin{gather}
{\displaystyle u_{xy} = \frac{\mu(u)}{\beta'(u_x)\gamma'(u_y)}, \qquad v_{xy} = \exp v, \qquad v = \beta(u_x) + \gamma(u_y) + \alpha(u),}\label{zhiber131}
\end{gather}
where
\begin{gather*}
{\displaystyle \alpha_1 u_x + \frac{\beta_1}{\beta'} = \exp \beta, \qquad \frac{\beta''}{\beta'^2} = c u_x \beta' + c_1, \qquad c_1 = - 1 -c, \qquad c \beta_1 = - \alpha_1}, \nonumber \\
{\displaystyle \frac{\gamma''}{\gamma'^2} = c u_y \gamma' + c_2, \qquad \frac{\alpha' \mu}{\gamma'} - \frac{\mu^2}{\gamma'} (c u_y \gamma' + c_1 + c_2) + \mu' u_y = \exp(\alpha + \gamma) \beta,} \nonumber \\
{\displaystyle -\alpha_1 \exp(\alpha + \gamma) + \alpha'' u_y + \frac{\mu' - c \mu^2}{\gamma'} = 0.}
\end{gather*}

Now, consider case \eqref{zhiber122}. Equations \eqref{zhiber124} and \eqref{zhiber125} can be rewritten in the forms 
\[
\alpha_1 u_x + \frac{\beta_1}{\beta'(u_x)} = \exp(\beta + \gamma_1) + 1, \qquad \mu_i \left( \alpha_1 u_x + \frac{\beta_1}{\beta'(u_x)} \right) = \exp(\beta + \gamma_i) + 1.
\]
It is not hard to show that
\[
\exp \beta \left( \mu_i \exp \gamma_1 - \exp \gamma_i \right) + \mu_i - 1 = 0.
\]
The dependence of $\beta$ only on the variable $u_x$ implies that $\mu_i = 1$ and $\gamma_i = \gamma_1$ for every $i$. This gives $\alpha(u) + \gamma(u_y) = \gamma_1$, where $\gamma_1$ is a constant, which contradicts the assumption $\gamma_{u_y} \neq 0$.

It remains to consider the case when $F$ is given by \eqref{zhiber123} to complete the analysis in the case when 
$(\alpha_i, \beta_i)$ are linearly dependent vectors. Using \eqref{zhiber123} we transform equations~\eqref{zhiber124} and~\eqref{zhiber125} into 
\begin{gather*}
\alpha_1 u_x + \frac{\beta_1}{\beta'(u_x)} = \exp(\beta + \gamma_1) + \exp (\sigma(\beta + \gamma_1)),\\
\mu_i \left( \alpha_1 u_x + \frac{\beta_1}{\beta'(u_x)} \right) = \exp(\beta + \gamma_i) + \exp (\sigma (\beta + \gamma_i)).
\end{gather*}
Consequently, we get
\[
\exp \beta \left( \mu_i \exp \gamma_1 - \exp \gamma_i \right) + \exp(\sigma \beta) \left( \mu_i \exp(\sigma \gamma_1) - \exp(\sigma \gamma_i) \right) = 0.
\]
Recall that $\sigma \neq 1$. Collecting coef\/f\/icients at $\exp \beta$ and $\exp( \sigma \beta) $ yields
\[
\mu_i \exp \gamma_1 = \exp \gamma_i, \qquad \mu_i \exp(\sigma \gamma_1) = \exp (\sigma \gamma_i).
\]
The above equations provide $\mu_i \exp(\sigma \gamma_1) (\mu^{\sigma - 1}_i - 1) = 0$, hence $\mu_i = 1$. It follows that $\gamma_i = \gamma_1$ for every $i$. By \eqref{zhiber124} we f\/ind that $\alpha(u) + \gamma(u_y) = \gamma_1$. This equation contradicts $\gamma_{u_y} \neq 0$.

Now, we must deal with the case when $\alpha_i$, $\beta_i$, $i=1,2$, satisfying $\alpha_1 \beta_2 - \beta_1 \alpha_2 \neq 0$ exist.
Setting def\/inite values of $u$, $u_y$ in $\eqref{zhiber114}$ we obtain the system
\[
\alpha_1 u_x + \frac{\beta_1}{\beta'(u_x)} = F(\beta(u_x) + \gamma_1), \qquad \alpha_2 u_x + \frac{\beta_2}{\beta'(u_x)} = F (\beta(u_x) + \gamma_2).
\]
Because of the given assumption $(u_x \beta')_{u_x} \neq 0$ we get
\begin{gather}
\kappa_1 F(\beta + \gamma_1) - \kappa_2 F(\beta + \gamma_2) = u_x, \qquad \kappa_3 F(\beta + \gamma_1) - \kappa_4 F(\beta + \gamma_2) = \frac{1}{\beta'}.  \label{zhiber132}
\end{gather}
We use
\begin{gather*}
\kappa_1 = \frac{\beta_2}{\alpha_1 \beta_2\! - \alpha_2 \beta_1}, \qquad \kappa_2 = \frac{\beta_1}{\alpha_1 \beta_2\! - \alpha_2 \beta_1}, \qquad \kappa_3 = \frac{\alpha_2}{\beta_1 \alpha_2\! - \beta_2 \alpha_1}, \qquad \kappa_4 = \frac{\alpha_1}{\beta_1 \alpha_2\! - \beta_2 \alpha_1}.
\end{gather*}

Let us analyze equation~\eqref{zhiber132} taking into account conditions \eqref{zhiber117}--\eqref{zhiber123}.

Consider the case when $F$ is given by \eqref{zhiber117}. It is not hard to show that equation~\eqref{zhiber132} implies $u_x = 0$. Thus, this case is not realized. Next, based on \eqref{zhiber118} we obtain that $u_x$ is a~constant. So it is also not possible.

If \eqref{zhiber119} is true then system \eqref{zhiber132} can be written as follows
\[
\kappa_1 (\beta + \gamma_1) - \kappa_2 (\beta + \gamma_2) = u_x, \qquad \kappa_3 (\beta + \gamma_1) - \kappa_4 (\beta + \gamma_2) = \frac{1}{\beta'}.
\]
It is not hard to verify that
\[
\beta'(\kappa_1 - \kappa_2) = 1, \qquad \beta (\kappa_3 - \kappa_4) + \gamma_1 \kappa_3 - \gamma_2 \kappa_4 = \kappa_1 - \kappa_2.
\]
Note that we used the properties $\kappa_1 - \kappa_2 \neq 0$, $\kappa_1 - \kappa_2 \neq 0$, which result from $\alpha_1 \beta_2 - \beta_1 \alpha_2 \neq 0$. Further, since $\kappa_3 - \kappa_4 \neq 0$, $\beta$ is a constant. This contradicts $\beta_{u_x} \neq 0$.

Let us discuss the case when the function $F$ is def\/ined by \eqref{zhiber120}. Rewriting \eqref{zhiber132} we get
\begin{gather*}
\kappa_1 (\beta + \gamma_1) \exp(\beta + \gamma_1) - \kappa_2 (\beta + \gamma_2) \exp(\beta + \gamma_2) = u_x, \\
\kappa_3(\beta + \gamma_1) \exp(\beta + \gamma_1) - \kappa_4 (\beta + \gamma_2) \exp(\beta + \gamma_2) = \frac{1}{\beta'}.
\end{gather*}
Setting $A = \kappa_1 \exp \gamma_1 - \kappa_2 \exp \gamma_2$ and $B = \kappa_1 \gamma_1 \exp \gamma_1 - \kappa_2 \gamma_2 \exp \gamma_2$ we obtain
\begin{gather}
u_x = A \beta \exp \beta + B \exp \beta.  \label{zhiber133}
\end{gather}
It is not dif\/f\/icult to determine that equations~\eqref{zhiber132}, \eqref{zhiber133} lead to
\begin{gather*}
(A\!+\!B)\!\left( \frac{\alpha' \mu}{\gamma'} - \frac{\mu^2}{\gamma'}\big(c_1\!+\!c u_y \gamma'\!+\!c_2\big) + \mu' u_y \right) +B\!\left( \alpha'' u_y + \frac{\mu'\!-\!c\mu^2}{\gamma'} \right) = (\alpha\!+\!\gamma)\exp(\alpha + \gamma),
\\
A \left( \frac{\alpha' \mu}{\gamma'} - \frac{\mu^2}{\gamma'} \big(c_1 + c u_y \gamma' + c_2\big) + \mu' u_y + \alpha'' u_y + \frac{\mu' - c \mu^2}{\gamma'} \right) = \exp(\alpha + \gamma).
\end{gather*}
Rewriting 
\eqref{zhiber113} by using \eqref{zhiber133} we f\/ind that $c = 1$, $c_1 = -2$. Thus, we obtain the equations
\begin{gather} \label{zhiber134}
{\displaystyle u_{xy} = \frac{\mu(u)}{\beta'(u_x)\gamma'(u_y)}, \qquad v_{xy} = v \exp v, \qquad v = \alpha(u) + \beta(u_x) + \gamma(u_y),}
\end{gather}
herewith
\begin{gather*}
{(A\!+\!B)\!\left( \frac{\alpha' \mu}{\gamma'} - \frac{\mu^2}{\gamma'}(u_y \gamma' -2 + c_2) + \mu' u_y \right) + B\!\left( \alpha'' u_y + \frac{\mu' - \mu^2}{\gamma'} \right) = (\alpha + \gamma) \exp(\alpha + \gamma),} \nonumber \\
{A \left( \frac{\alpha' \mu}{\gamma'} - \frac{\mu^2}{\gamma'} (u_y \gamma' -2 + c_2) + \mu' u_y + \alpha'' u_y + \frac{\mu' - \mu^2}{\gamma'}\right) = \exp(\alpha + \gamma),} \nonumber \\
{u_x = A \beta \exp \beta + B \exp \beta, \qquad \frac{\beta''}{\beta'^2} = u_x \beta' -2,\frac{\gamma''}{\gamma'^2} = u_y \gamma' + c_2.}
\end{gather*}
Note that case \eqref{zhiber121} yields the equations 
\eqref{zhiber131}.

Next, assume that the function $F$ is def\/ined by \eqref{zhiber122}. Hence, we write \eqref{zhiber132} as 
\begin{gather*}
\kappa_1 \left( \exp(\beta + \gamma_1) + 1 \right) - \kappa_2 \left( \exp(\beta + \gamma_2) + 1 \right) = u_x,\\
\kappa_3 \left( \exp(\beta + \gamma_1) +1 \right) - \kappa_4 \left( \exp(\beta + \gamma_2) + 1 \right) = \frac{1}{\beta'}.
\end{gather*}
Eliminating $\beta'$ from the last equation we get
\[
\exp \beta(\kappa_3 \exp \gamma_1 - \kappa_4 \exp \gamma_2 - \kappa_1 \exp \gamma_1 + \kappa_2 \exp \gamma_2) + \kappa_3 - \kappa_4 = 0.
\]
It is easy to show from this equation that $\beta$ is a constant. This contradicts $\beta_{u_x} \neq 0$.

Assuming that \eqref{zhiber123} holds, we can write \eqref{zhiber132} as
\begin{gather}
\exp \beta(\kappa_1 \exp \gamma_1 - \kappa_2 \exp \gamma_2) + \exp (\sigma \beta) (\kappa_1 \exp( \sigma \gamma_1) - \kappa_2 \exp (\sigma \gamma_2)) = u_x,
\nonumber \\
 \exp \beta (\kappa_3 \exp \gamma_1 - \kappa_4 \exp \gamma_2) + \exp (\sigma \beta) (\kappa_3 \exp (\sigma \gamma_1) - \kappa_4 \exp (\sigma \gamma_2)) = \frac{1}{\beta'}.\label{zhiber135}
\end{gather}
And further, from \eqref{zhiber113} based on \eqref{zhiber135} we obtain
\begin{gather}
(1 + c + c_1)(\kappa_1 \exp \gamma_1 - \kappa_2 \exp \gamma_2) \exp \beta \nonumber\\
\qquad{}+\big(\sigma^2 + c + c_1 \sigma\big)\big(\kappa_1 \exp (\sigma \gamma_1) - \kappa_2 \exp (\sigma \gamma_2)\big) \exp \sigma \beta = 0.\label{zhiber136}
\end{gather}
From \eqref{zhiber114} using \eqref{zhiber135} again we get
\begin{gather}
 (\kappa_1 \exp \gamma_1 - \kappa_2 \exp \gamma_2) \left( \frac{\alpha' \mu}{\gamma'} - \frac{\mu^2}{\gamma'}(c u_y \gamma' + c_1 + c_2) \right. \nonumber\\
\left.\qquad{} + \mu' u_y + \frac{\mu' - c \mu^2}{\gamma'} + \alpha'' u_y \right) = \exp(\alpha + \gamma),\label{zhiber137_0}
\\
(\kappa_1 \exp \sigma \gamma_1 - \kappa_2 \exp \sigma \gamma_2) \left( \sigma \left( \frac{\alpha' \mu}{\gamma'} - \frac{\mu^2}{\gamma'}(c u_y \gamma' + c_1 + c_2) + \mu' u_y \right) \right.\nonumber\\
\left. \qquad{}+  \frac{\mu' - c \mu^2}{\gamma'} + \alpha'' u_y \right) = \exp\sigma(\alpha + \gamma). \label{zhiber137_1}
\end{gather}
Note that if $\kappa_1 \exp (\sigma \gamma_1) - \kappa_2 \exp (\sigma \gamma_2) = 0$ then equations~\eqref{zhiber137_0} and \eqref{zhiber137_1} imply that $\exp \sigma(\alpha + \gamma) = 0$. Consequently, the equalities
$1+ c + c_1 = 0$ and $\sigma^2 + c_1 \sigma + c = 0$ arise from equation \eqref{zhiber136}.
The solution of the last equation is found as $\sigma = c$, where $c = -1 - c_1$. Thus, denoting $A = \kappa_1 \exp \gamma_1 - \kappa_2 \exp \gamma_2$, $B = \kappa_1 \exp (\sigma \gamma_1) - \kappa_2 \exp( \sigma \gamma_2)$ we obtain
\begin{gather} \label{zhiber138}
{\displaystyle u_{xy} = \frac{\mu(u)}{\beta'(u_x) \gamma'(u_y)}, \qquad v_{xy} = \exp v + \exp (\sigma v), \qquad v = \alpha(u) + \beta(u_x) + \gamma(u_y),}
\end{gather}
where
\begin{gather*}
 A \exp \beta + B \exp (\sigma \beta) = u_x, \qquad \frac{\beta''}{\beta'^2} = \sigma u_x \beta' -1-\sigma, \qquad \frac{\gamma''}{\gamma'^2} = \sigma u_y \gamma' + c_2 , \nonumber \\
 A \left( \frac{\alpha' \mu}{\gamma'} - \frac{\mu^2}{\gamma'}\big(\sigma u_y \gamma'+c_2-1\big) + \mu' u_y + \frac{\mu' }{\gamma'} + \alpha'' u_y \right) = \exp(\alpha + \gamma), \nonumber \\
 B \left( \sigma \left( \frac{\alpha' \mu}{\gamma'} - \frac{\mu^2}{\gamma'}\big(\sigma u_y \gamma' + c_2 - \sigma\big) + \mu' u_y \right) +
 \frac{\mu'}{\gamma'} + \alpha'' u_y \right) = \exp\sigma(\alpha + \gamma).
\end{gather*}

Let us discuss the results obtained. We should analyze the equations and conditions for the parameters found in cases \eqref{zhiber117}--\eqref{zhiber123} and use the fact that functions \eqref{zhiber19} and \eqref{zhiber22} are invariant under the permutation of $\beta(u_x)$ and $\gamma(u_y)$.

In case \eqref{zhiber121} we obtained \eqref{zhiber131}. By interchanging $\beta(u_x)$ and $\gamma(u_y)$ we get
\begin{gather}
 \alpha_2 u_y + \frac{\beta_2}{\gamma'} = \exp \gamma, \qquad \frac{\gamma''}{\gamma'^2} = c u_y \gamma' + c_2, \qquad c_2 = -1-c, \qquad c \beta_2 = - \alpha_2,\nonumber\\
\frac{\alpha' \mu}{\beta'} - \frac{\mu^2}{\beta'} (c u_x \beta' + c_1 + c_2) + \mu' u_x = \exp (\alpha + \beta) \beta_2, \nonumber\\
 -\alpha_2 \exp(\alpha+ \beta) + \alpha'' u_x + \frac{\mu' - c \mu^2}{\beta'} = 0. \label{zhiber139}
\end{gather}
We substitute $\gamma$ satisfying the conditions for the parameters listed for equation~\eqref{zhiber139} in\-to~\eqref{zhiber131}. At the same time we substitute $\beta$ satisfying the conditions for the parameters listed for equation~\eqref{zhiber131} into \eqref{zhiber139}. As a result, we obtain the system
\begin{gather*}
\frac{1}{\beta_2} (\exp \gamma - \alpha_2 u_y)\big(\alpha' \mu + 2 \mu^2 (1+c)\big) + \mu' u_y = \exp(\alpha+ \gamma) \beta_1,\\
-\alpha_1 \exp(\alpha+\gamma) + \alpha'' u_y + \big(\mu' - c \mu^2\big) \frac{1}{\beta_2}(\exp \gamma - \alpha_2 u_y) = 0,\\
\frac{1}{\beta_1}(\exp \beta - \alpha_1 u_x)\big(\alpha' \mu + 2 \mu^2(1+c)\big) + \mu' u_x = \exp(\alpha + \beta) \beta_2,\\
- \alpha_2 \exp(\alpha+ \beta) + \alpha'' u_x + \big(\mu' - c \mu^2\big) \frac{1}{\beta_1}(\exp \beta - \alpha_1 u_x) = 0.
\end{gather*}
Since $\exp \gamma$ and $u_y$, $\exp \beta$ and $u_x$ are independent, equations~\eqref{u_xy}, \eqref{v_xy}, and \eqref{phi} take the following forms:
\[
u_{xy} = \frac{\mu(u)}{\beta'(u_x)\gamma'(u_y)}, \qquad v_{xy} = \exp v, \qquad v = \alpha(u) + \beta(u_x) + \gamma(u_y),
\]
where $\alpha$ and $\beta$ are solutions of the ordinary dif\/ferential equations
\[\alpha_1 u_x + \frac{\beta_1}{\beta'} = \exp \beta, \qquad \alpha_2 u_y + \frac{\beta_2}{\gamma'(u_y)} = \exp \gamma, \qquad -\frac{\alpha_1}{\beta_1} = - \frac{\alpha_2}{\beta_2} = c,
\]
and the functions $\mu$ and $\alpha$ satisfy
\begin{gather*}
\alpha' \mu + 2 \mu^2 (c + 1) = \beta_1 \beta_2 \exp \alpha, \qquad c \beta_1 \beta_2 \exp \alpha + \mu' - c \mu^2 = 0, \qquad \alpha'' + c(\mu' - c \mu^2) = 0.
\end{gather*}
Analyzing the last system we obtain cases \eqref{zhiber109}, \eqref{zhiber110}. It is easy to verify that case \eqref{zhiber120} is not possible.

Based on \eqref{zhiber123} we get \eqref{zhiber138}. Interchanging $\beta(u_x)$ and $\gamma(u_y)$ implies
\begin{gather} \label{zhiber141}
\begin{split}
&{A_2 \exp \gamma + B_2 \exp \sigma \gamma = u_y, \qquad \frac{\gamma''}{\gamma'^2} = \sigma u_y \gamma' + c_2, \qquad c_2 = -1-c,}\\
&{\displaystyle \frac{A_2}{\beta'}\left( \alpha' \mu - \mu^2(c_1 - 1) + \mu' \right) + A_2 u_x \left( -\sigma \mu^2 + \mu' + \alpha'' \right) = \exp(\alpha + \beta),}\\
&{\displaystyle \frac{B_2}{\beta'}\left( \sigma(\alpha' \mu - \mu^2 (c_2 - \sigma)) + \mu' \right) +u_x B_2 \left( \sigma( -\sigma \mu^2 + \mu') + \alpha'' \right) = \exp\sigma(\alpha+\beta).}
\end{split}
\end{gather}
Similarly, we substitute $\beta$ satisfying the conditions for the parameters listed for equation~\eqref{zhiber138} into \eqref{zhiber141} and obtain
\begin{gather*}
 (A \exp \beta + B \sigma \exp \sigma \beta)A_2(\alpha' \mu + \mu^2(2+\sigma) + \mu') \\
\qquad{} +(A \exp \beta + B \exp \sigma \beta)A_2(\mu' - \sigma \mu^2 + \alpha'') = \exp(\alpha + \beta),
\\
(A \exp \beta + B \sigma \exp(\sigma \beta))B_2 \left( \sigma(\alpha' \mu + \mu^2(1 + 2\sigma) + \mu') \right) \\
\qquad{} +(A \exp \beta + B \exp \sigma \beta) B_2 \left( \sigma(\mu' - \sigma \mu^2) + \alpha'' \right) = \exp \sigma(\alpha + \beta).
\end{gather*}
Taking into account the fact that $\exp \beta$, $\exp \sigma \beta$ are independent, we get
\begin{gather*}
A A_2 \left( \alpha' \mu + \mu^2(2 + \sigma) + 2 \mu' - \sigma \mu^2 + \alpha'' \right) = \exp \alpha, \\
\sigma \alpha' \mu + \sigma (\sigma + 1) \mu^2 + (\sigma + 1)\mu' + \alpha'' = 0, \\
B B_2 \left( \sigma^2 \alpha' \mu + 2 \sigma^3 \mu^2 + 2 \sigma \mu' + \alpha'' \right) = \exp (\sigma \alpha).
\end{gather*}
Solving the above system we obtain cases \eqref{zhiber111} and \eqref{zhiber112}.
\end{proof}

\subsection[Case $\varphi = c \ln u_x + q(u, u_y)$]{Case $\boldsymbol{\varphi = c \ln u_x + q(u, u_y)}$}

We have the following statement in this case.
\begin{lemma}\label{lemma6}
Suppose that \eqref{zhiber20} is satisfied. Then equations~\eqref{u_xy}, \eqref{v_xy}, and \eqref{phi} take the following forms
\begin{gather} \label{zhiber143}
u_{xy} = \frac{\mu(u) - q_u(u,u_y)}{q_{u_y}(u,u_y)}u_x, \qquad v_{xy} = c_2 \exp v, \qquad v = \ln u_x + q(u,u_y),
\end{gather}
where
\[
 \frac{\mu - q_u}{q_{u_y}}\left( \mu - \frac{\mu - q_u}{q^2_{u_y}} q_{u_y u_y} - 2\frac{q_{u u_y}}{q_{u_y}} \right) + \frac{\mu'}{q_{u_y}} - \frac{q_{uu}}{q_{u_y}} + \mu' u_y = c_2 \exp q, \qquad q_{uu_y} \neq 0,
\]
up to the point transformations $u \rightarrow \theta(u)$, $v \rightarrow \kappa(v)$, $x \rightarrow \xi x$, and $y \rightarrow \eta y$, where $\xi$ and $\eta$ are arbitrary constants.
\end{lemma}

\begin{proof}
Substituting function \eqref{zhiber20} into equation~\eqref{zhiber12} we obtain
\[
A(u,u_y) q_{u_y}(u,u_y) - q_u(u, u_y) q_{u_y}(u, u_y) u_y + c q_u(u, u_y) = B(u,u_x)\frac{c}{u_x}.
\]
Recall that $u_x$, $u_y$ are considered as independent variables. Hence, the above equation is equivalent to the system
\[
\displaystyle A q_{u_y} - q_u q_{u_y} u_y + c q_u = \mu(u), \qquad \frac{B c}{u_x} = \mu(u).
\]
From these equations we f\/ind the functions $A$ and $B$, 
\[
B = \frac{\mu u_x}{c}, \qquad A = \frac{\mu + q_u q_{u_y} u_y - c q_u}{q_{u_y}}.
\]
 By using these equations in each of equations~\eqref{secondcondition}, \eqref{thirdcondition} we determine the function $f$ of equation~\eqref{u_xy} as
\[
f = \frac{\mu - c q_u}{c q_{u_y}} u_x.
\]
Substituting the functions \eqref{zhiber20} and $f$ into \eqref{zhiber13} we have
\[
u_x \left(
\frac{\mu - cq_u}{cq_{u_y}}\left( \frac{\mu}{c} - \frac{\mu - cq_u}{q^2_{u_y}} q_{u_y u_y} - 2c\frac{q_{u u_y}}{q_{u_y}} \right) + \frac{\mu'}{q_{u_y}} - \frac{q_{uu}}{q_{u_y}} + \frac{\mu' u_y}{c} \right) = F(c \ln u_x + q).
\]
It is not dif\/f\/icult to prove by dif\/ferentiating this equation with respect to $u_x$ that $c F' = F$. Consequently,
$
F(z) = c_2 \exp(z/c).
$
Here $c_2$ is an arbitrary constant. Thus, equations~\eqref{u_xy}, \eqref{v_xy}, and \eqref{phi} are of the forms
\[
u_{xy} = \frac{\mu(u) - c q_u(u,u_y)}{c q_{u_y}(u,u_y)} u_x, \qquad v_{xy} = c_2 \exp(v/c), \qquad v = c \ln u_x + q(u,u_y),
\]
where
\[
\frac{\mu - cq_u}{cq_{u_y}}\left( \frac{\mu}{c} - \frac{\mu - cq_u}{q^2_{u_y}} q_{u_y u_y} - 2c\frac{q_{u u_y}}{q_{u_y}} \right ) + \frac{\mu'}{q_{u_y}} - c\frac{q_{uu}}{q_{u_y}} + \frac{\mu' u_y}{c} = c_2 \exp (q/c).
\]
Finally, the transformations $v \rightarrow c v$, $q \rightarrow c q$, $\mu \rightarrow c^2 \mu$, and $c_2 / c \rightarrow c_2$ transform these equations into \eqref{zhiber143}.
\end{proof}

\subsection[Case $\varphi = \alpha(u) + \kappa(u) \ln u_x + \mu(u) \ln u_y$]{Case $\boldsymbol{\varphi = \alpha(u) + \kappa(u) \ln u_x + \mu(u) \ln u_y}$}

By substituting \eqref{zhiber21} into \eqref{zhiber12} we obtain
\begin{gather*}
\left( A(u,u_y) -(\kappa'(u)\ln u_x + \mu'(u) \ln u_y + \alpha'(u)) u_y \right) \frac{\mu(u)}{u_y} \\
\qquad{} = \left( B(u, u_y) - (\kappa'(u)\ln u_x + \mu'(u) \ln u_y + \alpha'(u)) u_x \right) \frac{\kappa(u)}{u_x},
\end{gather*}
which can be written as 
\begin{gather*}
\frac{B(u,u_x) \kappa(u)}{u_x} + \left( \kappa'(u) \ln u_x + \alpha'(u) \right) \left( \mu(u) - \kappa(u) \right) \\
\qquad{}= \frac{A(u, u_y) \mu(u)}{u_y} - \mu'(u) \ln u_y \left( \mu(u) - \kappa(u) \right).
\end{gather*}
Since $u_x$ and $u_y$ are regarded as independent variables, the above equation is equivalent to the system
\begin{gather*}
\frac{B(u,u_x) \kappa(u)}{u_x} + \left( \kappa'(u) \ln u_x + \alpha'(u) \right) \left( \mu(u) - \kappa(u) \right) = \lambda(u), \\
\frac{A(u, u_y) \mu(u)}{u_y} - \mu'(u) \ln u_y \left( \mu(u) - \kappa(u) \right) = \lambda(u).
\end{gather*}
The formulae
\[
B = \frac{\left( \lambda - (\mu - \kappa)(\kappa' \ln u_x + \alpha') \right)u_x}{\kappa}, \qquad A= \frac{\left( \lambda + \mu' (\mu - \kappa) \ln u_y \right) u_y}{\mu}
\]
thereby immediately follow. Substituting $A$ and $B$ into equations~\eqref{secondcondition} and \eqref{thirdcondition} we f\/ind $f$, 
\begin{gather}
f = \frac{\lambda - \kappa \mu' \ln u_y - \mu \kappa' \ln u_x - \mu \alpha'}{\kappa \mu} u_x u_y. \label{zhiber147}
\end{gather}
We apply the operator $\frac{\partial}{\partial u_x}$ to both sides of equation~\eqref{fourthcondition} and use the equations obtained. So we get $F' \kappa = F$, while applying $\frac{\partial}{\partial u_y}$ implies $F' \mu = F$. This requires $\mu(u) = \kappa(u) = c$. Thus~$\varphi$ takes the form $\varphi = \alpha(u) + c \ln( u_x u_y)$, and case \eqref{zhiber21} is reduced to case \eqref{zhiber19} considered earlier.

Theorem~\ref{theorem1} follows from Lemmas~\ref{lemma1}--\ref{lemma6}.

\section[Differential substitutions of the form $u=\psi(v,v_x,v_y)$]{Dif\/ferential substitutions of the form $\boldsymbol{u=\psi(v,v_x,v_y)}$}
\label{section4}

In this section we consider the problem which is, in a sense, inverse to the original problem. The aim is to describe equations of form \eqref{v_xy} which are transformed into equations of form~\eqref{u_xy} by dif\/ferential substitutions~\eqref{kuzn_psi}.

\begin{theorem}\label{theorem2}
Suppose that equation~\eqref{v_xy} is transformed into equation~\eqref{u_xy} by differential substitution~\eqref{kuzn_psi}. Then equations~\eqref{v_xy}, \eqref{u_xy} and substitution~\eqref{kuzn_psi} take one of the following forms:
\begin{alignat*}{6}
& v_{xy} = v, \qquad && u_{xy} = u, \qquad && u = c_1 u_x + c_2 u_y + c_3 u; & \\
& v_{xy} = 0, \qquad && u_{xy} = 0, \qquad && u = \beta(v_x) + \gamma(v_y) + c_3 v; &\\
& v_{xy} = 0, \qquad && u_{xy} = \exp(u) u_y, \qquad && u = \ln \left( - \frac{p'(v) v_x}{\mu(v_y) + p(v)} \right), &
\intertext{where $p'(v) = \exp(cv)$;}
& v_{xy} = 1, \qquad && u_{xy} = c_1 (u_x - c_2), \qquad && u = \exp(c_1 v_x) + c_2 v_y; & \\
& v_{xy} = \exp v, \qquad && u_{xy} = u u_x, \qquad && u = v_y + \mu(v_x) \exp v, &
\intertext{where $2 \mu' = \mu^2$;}
& v_{xy} = 0, \qquad && u_{xy} = \exp u, \qquad && u = \ln (v_x v_y) + \delta(v), &
\intertext{where $\delta''(v) = \exp \delta(v)$;}
& v_{xy} = 1, \qquad && u_{xy} = c_1 u_x + c_2 u_y - c_1 c_2 u, \qquad && u = \exp(c_1 v_x) + \exp(c_2 v_y) &
\end{alignat*}
up to the point transformations $u \rightarrow \theta(u)$, $v \rightarrow \kappa(v)$, $x \rightarrow \xi x$, and $y \rightarrow \eta y$ and the substitution 
$u + \xi x + \eta y \rightarrow u$, where $\xi$ and $\eta$ are arbitrary constants. Here $c$ is an arbitrary constant, $c_1$~and~$c_2$ are nonzero constants.
\end{theorem}

Note that symmetries, $x$- and $y$-integrals, and the general solutions of the equations $u_{xy} = u u_x$ and $u_{xy} = \exp(u) u_y$ were given in~\cite{MeshSok}.
The transformation connecting the Liouville equation to the wave equation is well known (see~\cite{ZhSok}).

Here we just give the outline of the proof.

\begin{proof}[Scheme of the proof.] Substituting the function $\psi$ given by \eqref{kuzn_psi} into equation~\eqref{u_xy} and \mbox{using}~\eqref{v_xy} we obtain
\begin{gather}
 \psi_v F + \psi_{v_x} F' v_x + \psi_{v_y} F' v_y + v_x \bigl( \psi_{vv} v_y + \psi_{v v_x} F + \psi_{v v_y} v_{yy} \bigr) \nonumber\\
\qquad{} +v_{xx} \bigl( \psi_{v_x v} v_y + \psi_{v_x v_x} F + \psi_{v_x v_y} v_{yy} \bigr) + \bigl( \psi_{v_y v} v_y + \psi_{v_y v_x} F + \psi_{v_y v_y} v_{yy}\bigr) F \nonumber\\
\qquad\quad{} = f \bigl( \psi, \psi_v v_x + \psi_{v_x} v_{xx} + \psi_{v_y}F, \psi_v v_y + \psi_{v_x} F + \psi_{v_y} v_{yy} \bigr).\label{zhiber_eq4_8}
\end{gather}
Denote the arguments of the function $f$ by $a$, $b$, and $c$. Recall that we have $\psi_{v_x} \psi_{v_y} \neq 0$. The equality $f''_{bb} = f''_{cc} = 0$ thereby immediately follows from equation~\eqref{zhiber_eq4_8}. Hence, equation~\eqref{u_xy} takes the form
\[
u_{xy} = \alpha(u) + \beta(u) u_x + \gamma(u) u_y + \epsilon(u) u_x u_y.
\]
After the point transformation $u \rightarrow A(u)$ with $A'' - \epsilon A'^2 = 0$ the above equation takes the form
\[
u_{xy} = f = \alpha(u) + \beta(u) u_x + \gamma(u) u_y.
\]
Next, taking into account the last equality which def\/ines the function f we can rewrite equation~\eqref{zhiber_eq4_8} as follows
\begin{gather*}
 \psi_v F + \psi_{v_x} F' v_x + \psi_{v_y} F' v_y + v_x \bigl( \psi_{vv} v_y + \psi_{v v_x} F + \psi_{v v_y} v_{yy} \bigr) \\
\qquad{} + v_{xx} \bigl( \psi_{v_x v} v_y + \psi_{v_x v_x} F + \psi_{v_x v_y} v_{yy} \bigr) + \bigl( \psi_{v_y v} v_y + \psi_{v_y v_x} F + \psi_{v_y v_y} v_{yy}\bigr) F \\
\qquad\quad{} = \alpha(\psi) + \beta(\psi) \bigl( \psi_v v_x + \psi_{v_x} v_{xx} + \psi_{v_y}F \bigr) + \gamma(\psi) \bigl( \psi_v v_y + \psi_{v_x} F + \psi_{v_y} v_{yy} \bigr).
\end{gather*}
Since $v_{xx}$ and $v_{yy}$ are independent variables, this equation is equivalent to the system
\begin{gather*}
\psi_{v_x v_y} = 0, \\
\psi_{v_x v} v_y + \psi_{v_x v_x} F = \beta(\psi) \psi_{v_x}, \\
\psi_{v_y v} v_x + F \psi_{v_y v_y} = \gamma(\psi) \psi_{v_y}, \\
\psi_v F + \psi_{v_x} F' v_x + \psi_{v_y} F' v_y + \psi_{vv} v_x v_y + v_x \psi_{v v_x} F + v_y \psi_{v v_y} F + F^2 \psi_{v_y v_x} \\
\qquad{}= \alpha(\psi) + \beta(\psi) \bigl( \psi_v v_x + \psi_{v_x} v_{xx} + \psi_{v_y}F \bigr) + \gamma(\psi) \bigl( \psi_v v_y + \psi_{v_x} F + \psi_{v_y} v_{yy} \bigr).
\end{gather*}
Consequently, we have
\begin{gather*}
\psi = A(v, v_x) + B(v, v_y), \\
A_{v v_x} v_y + A_{v_x v_x} F = \beta(A+B) A_{v_x}, \\
B_{v v_y} v_x + B_{v_y v_y} F = \gamma(A+B) B_{v_y}, \\
(A_v + B_v) F + A_{v_x} F' v_x + B_{v_y} F' v_y + (A_{vv} + B_{vv}) v_x v_y + v_x A_{v v_x} F + v_y B_{v v_y} F \\
\qquad{}= \alpha(A+B) + \beta(A+B) \bigl( v_x(A_v + B_v) + F B_{v_y} \bigr) + \gamma(A+B) \bigl( v_y (A_v + B_v) + A_{v_x} F \bigr).
\end{gather*}
By using the above equations we prove Theorem~\ref{theorem2}.
\end{proof}

\subsection*{Acknowledgements}

This work is partially supported by the
Russian Foundation for Basic Research (RFBR) (Grants 11-01-97005-Povolj'ie-a, 12-01-31208 mol-a).

\pdfbookmark[1]{References}{ref}
\LastPageEnding


\begin{thebibliography}{99}
\footnotesize\itemsep=0pt

\bibitem{Anderson2}
Anderson I.M., Kamran N., The variational bicomplex for hyperbolic second-order
  scalar partial dif\/ferential equations in the plane, \href{http://dx.doi.org/10.1215/S0012-7094-97-08711-1}{\textit{Duke Math.~J}}
  \textbf{87} (1997), 265--319.

\bibitem{Backl1}
B\"acklund A.V., Einiges \"uber Curven und Fl\"achen Transformationen,
  \textit{Lund Universit\"ets Arsskrift} \textbf{10} (1874), 1--12.

\bibitem{Bianchi}
Bianchi L., Ricerche sulle superf\/icie elicoidali e sulle superf\/icie a curvatura
  costante, \textit{Ann. Scuola Norm. Sup. Pisa Cl. Sci.} \textbf{2} (1879),
  285--341.

\bibitem{Darb}
Darboux G., Le\c{c}ons sur la th\'eorie g\'en\'erale des surfaces et les
  applications g\'eom\'etriques du calcul inf\/init\'esimal.~II,
  Gauthier-Villars, Paris, 1889.

\bibitem{DrSvSok}
Drinfel'd V.G., Svinolupov S.I., Sokolov V.V., Classif\/ication of f\/ifth-order
  evolution equations having an inf\/inite series of conservation laws,
  \textit{Dokl. Akad. Nauk Ukrain. SSR Ser. A}  (1985), no.~10, 8--10.

\bibitem{Goursat}
Goursat E., Le\c{c}on sur l'int\'egration des \'equations aux d\'eriv\'ees
  partielles du second ordre \'a deux variables ind\'ependantes, I,~II,
  Hermann, Paris, 1896.

\bibitem{Kh2}
Khabirov S.V., Inf\/inite-parameter families of solutions of nonlinear
  dif\/ferential equations, \href{http://dx.doi.org/10.1070/SM1994v077n02ABEH003442}{\textit{Sb. Math.}} \textbf{77} (1994), 303--311.

\bibitem{Kuzn1}
Kuznetsova M.N., Laplace transformation and nonlinear hyperbolic equations,
  \textit{Ufa Math.~J.} \textbf{1} (2009), no.~3, 87--96.

\bibitem{Kuzn2}
Kuznetsova M.N., On nonlinear hyperbolic equations related with the
  Klein--Gordon equation by dif\/ferential substitutions, \textit{Ufa Math.~J.}
  \textbf{4} (2012), no.~3, 86--103.

\bibitem{Liouville}
Liouville J., Sur l'equation aux dif\/f\'erences partielles $\partial^2 \log
  \lambda /\partial u\partial v \pm \lambda /(aa^2)=0$, \textit{J.~Math. Pures
  Appl.} \textbf{18} (1853), 71--72.

\bibitem{MeshSok}
Meshkov A.G., Sokolov V.V., Hyperbolic equations with third-order symmetries,
  \href{http://dx.doi.org/10.1007/s11232-011-0004-3}{\textit{Theoret. Math. Phys.}} \textbf{166} (2011), 43--57.

\bibitem{Sok}
Sokolov V.V., On the symmetries of evolution equations, \href{http://dx.doi.org/10.1070/RM1988v043n05ABEH001927}{\textit{Russian Math. Surveys}}
  \textbf{43} (1988), no.~5, 165--204.

\bibitem{SolAbdo}
Soliman A.A., Abdo H.A., New exact solutions of nonlinear variants of the RLN,
  the PHI-four and Boussinesq equations based on modif\/ied extended direct
  algebraic method, \textit{Int.~J. Nonlinear Sci.} \textbf{7} (2009),
  274--282, \href{http://arxiv.org/abs/1207.5127}{arXiv:1207.5127}.

\bibitem{St2}
Startsev S.Ya., Hyperbolic equations admitting dif\/ferential substitutions,
  \href{http://dx.doi.org/10.1023/A:1010359808044}{\textit{Theoret. Math. Phys.}} \textbf{127} (2001), 460--470.

\bibitem{St1}
Startsev S.Ya., Laplace invariants of hyperbolic equations linearizable by a
  dif\/ferential substitution, \href{http://dx.doi.org/10.1007/BF02557408}{\textit{Theoret. Math. Phys.}} \textbf{120} (1999),
  1009--1018.


\bibitem{Sv}
Svinolupov S.I., Second-order evolution equations with symmetries,
  \href{http://dx.doi.org/10.1070/RM1985v040n05ABEH003693}{\textit{Russian Math. Surveys}} \textbf{40} (1985), no.~5, 241--242.

\bibitem{Tzitz}
Tzitz\'eica G., Sur une nouvelle classe de surfaces, \textit{C.~R. Acad. Sci.}
  \textbf{144} (1907), 1257--1259.

\bibitem{ZhiberShabat}
Zhiber A.V., Shabat A.B., Klein--Gordon equations with a nontrivial group,
\textit{Soviet Phys. Dokl.} \textbf{24} (1979), 607--609.

\bibitem{ZhSok}
Zhiber A.V., Sokolov V.V., Exactly integrable hyperbolic equations of
  {L}iouville type, \href{http://dx.doi.org/10.1070/rm2001v056n01ABEH000357}{\textit{Russian Math. Surveys}} \textbf{56} (2001), no.~1,
  61--101.

\bibitem{ZhSokSt}
Zhiber A.V., Sokolov V.V., Startsev S.Ya., Darboux integrable nonlinear
  hyperbolic equations, \textit{Dokl. Math.} \textbf{52} (1995), 128--130.

\end{thebibliography}
\end{document}